\documentclass[twocolumn, twocolappendix]{aastex631}
\usepackage{amsmath}
\usepackage{multirow}
\usepackage{natbib}
\usepackage{polski}
\renewcommand{\figurename}{Figure}

\shorttitle{Magnetically driven relativistic outflow}
\shortauthors{Xie \& Lei}
\graphicspath{{./}{figures/}}

\begin{document}
\title{The relativistic outflow driven by the large-scale magnetic field from an accretion disk}

\author[0000-0001-5553-4577]{Wei Xie}
\affiliation{Department of Astronomy, School of Physics and Electronic Science, Guizhou Normal University, Guiyang 520001, China. Email: xieweispring@gznu.edu.cn}
\affiliation{Guizhou Provincial Key Laboratory of Radio Astronomy and Data Processing, Guizhou Normal University, Guiyang 520001, China.}
\author[0000-0003-3440-1526]{Wei-Hua Lei}
\affiliation{Department of Astronomy, School of Physics, Huazhong University of Science and Technology, Wuhan 430074, China. Email: leiwh@hust.edu.cn}




\begin{abstract}
Outflows/jets are ubiquitous in a wide range of astrophysical objects, yet the mechanisms responsible for their generation remain elusive. One hypothesis is that they are magnetically driven. Based on general relativistic MHD equations, we establish a formulation to describe the outflows driven by large-scale magnetic fields from the accretion disk in Schwarzschild spacetime. The outflow solution manifests as a contour level of a ``Bernoulli" function, which is determined by ensuring that it passes through both the slow and fast magnetosonic points. This approach is a general relativistic extension to the classical treatment of Cao and Spruit (1994). The initial plasma $\beta$ that permits magnetically driven outflow solutions is constrained, with the slow magnetosonic point above the footpoint setting an upper limit ($\beta_\mathrm{b}\lesssim 2$) and the Alfv\'en point inside the light cylinder setting a lower limit ($\beta_\mathrm{b}\gtrsim 0.02$). The higher the magnetization, the higher the temperature allowed, leading to relativistic outflows/jets. We investigate the relativistic outflows/jets of several typical objects such as active galactic nuclei (AGN), X-ray binaries (XRBs) and gamma-ray bursts (GRBs). The results indicate that all of these phenomena require strongly magnetized, high-temperature outflows as initial conditions, suggesting a potential association between the production of relativistic outflows/jets and corona-like structures.

\end{abstract}



\section{Introduction} \label{sec:intro}
Outflows/Jets are ubiquitous in astrophysical objects of various scales, including young stellar objects (YSOs), active galactic nuclei (AGN), X-ray binaries (XRBs), and gamma-ray bursts (GRBs). It is still unknown how these jets are produced. Nontheless, people believe that those different jet might share common production mechanism in which the combination of magnetic fields and rotation play an important role. The plasma acceleration process in the ordered magnetic field has been extensively explored. \cite{WD67} (hereafter WD67) developed the first quantitative theory of magnetic stellar wind. Their model deals with the steady, polytropic, axisymmetric flow of an inviscid and perfectly conducting fluid near the equator, and they found that such a wind can accelerate particles up to very high velocities and is efficient to carry off most of the angular momentum of the star. An important feature of magnetic winds is the existence of three ``critical points" where the flow velocity equals the wave velocity of three MHD wave modes (the slow, Alfv\'en and fast modes). This situation is more complicated than non-magnetic wind \citep{Parker58} which has only one ``critical point" where the flow velocity equals the sound speed. \cite{Yeh1976} applied the formulation of WD67 equally to the region outside the equator. The poloidal magnetic field is assumed in WD67, so it is referred to as a 1D model. To be more realistic, a series of later works made efforts to self-consistently solve the poloidal magnetic field by considering the force-balance across the field lines (e.g., \citealp{PK71, Sakurai85, Sakurai87}). An important conclusion of \citealt{Sakurai85} and \cite{Sakurai87} is that the field lines are deflected toward the direction of the rotation axis and the flow becomes collimated at large distances. The driver of the collimation is the toroidal magnetic field which develops in the wind due to the rotation of either the central object or the disk. In the case of accretion disks, the inclination of the filed lines at the disk surface has strong influence on the resulting magnetically driven outflow \citep{BP82}. \cite{CaoSpruit1994} (hereafter CS94) investigate in detail how the wind properties change as the field inclination at the disk surface. For field lines inclined less than $\sim 60^\circ$, the wind acceleration can start immediately at the disk surface and leads to a large mass flux and consequently a very low terminal speed and highly wound-up field lines. For more steeply inclined fields the flow first has to overcome a barrier of increasing effective potential before acceleration away from the disk. 

Above works are based on Newtonian magnetic hydrodynamics, the development from classical Newtonian theory to the theory of relativity follows a logical and natural progression. With the increasing discovery of ultra-fast outflows (UFOs) with velocities $v\gtrsim0.03c$ (even $v\gtrsim0.3c$) in AGN (\citealp{Reeves2009, Tombesi2010a, Tombesi2010b, Tombesi2011, Tombesi2013, Tombesi2015, Gofford2013, Gofford2015, Parker2017, Reeves2020, Matzeu2023, Lanzuisi2024}) and XRBs (\citealp{Miller2006, Miller2008, Miller2012, King2014, Kosec2018, Wang2021}), the magnetically driven outflow theory based on relativity has become a practical necessity. The first extension of the theory of WD67 to relativistic winds was done by \cite{Michel69} who was considering cold outflows driven by rapidly rotating highly magnetized neutron stars. \cite{GJ70} explored cool isothermal relativistic winds. \cite{Kennel83} extended Michel's model to finite temperatures and relativistic injection speeds. All these works were limited to the equatorial plane and either completely neglected the effect of gravity or adopted an approximate treatment for it. \cite{Okamoto78} firstly derived an exact general relativistic description for magnetic winds beyond the equatorial plane. However, this work was restricted to pressureless flows only. In a series of papers \cite{Camenzind1986, Camenzind1986b, Camenzind1987} derived a complete set of equations describing a stationary axisymmetric relativistic magnetic wind in an arbitrary metric. He then solved these equations in cold flows and studied the jet geometries. \cite{Cao1997} dealt with the centrifugal acceleration of the cold gas in a Kerr metric by assuming the gas particles to be beads threaded by field lines, it is found that jets can be centrifugally launched even for nearly vertically shaped magnetic flux surfaces as long as the black hole is fast rotating. This result is confirmed by \cite{SS2010} in a more careful calculation. \cite{DaigneDrenkhahn2002} made efforts to extend the model of WD67 with including an exact treatment of all effects (thermal pressure, gravity and arbitrary shapes of flux tubes). \cite{Li03} self-consistently combined the dynamics of the magnetic wind and the accretion process around a Kerr black hole. However, both of their works were limited to the region near the equatorial plane. \cite{Pu2015} present a semi-analytical GRMHD solution for the cold inflow/outflow along large-scale magnetic fields threading a rotating black hole, which can be considered as a more realistic scenario of Blandford-Znajek process \citep{BZ77}. The outflows driven by the large-scale magnetic field threading the surface of either the central object or the accretion disk is uniformally treated by \cite{CZ2021} under some analytic assumptions for the scaling rules of the magnetic field. Although this work encompasses the study of jets/outflows across the nonrelativistic to relativistic regimes, the thermal dynamics are still neglected, and the relativistic effects are approximately dealt with.

The goal of this work is to present a general relativistic formulation of the equations governing a thermal stationary axisymmetric MHD flow driven by the large-scale magnetic fields threading an accretion disk, as an relativistic extension to CS94, so that it can be directly applied to the ultra-fast outflows even the jets of the compact objects like AGN, GRBs, XRBs, etc. We describe the formulation of the model in Section \ref{sec:outflowEqs}. The numerical results of the model and some discussions about its possible implications in outflows of AGN/XRBs/GRBs are given in Section \ref{sec:results}, and we make a summary in Section \ref{sec:summary}.

\section{Model} \label{sec:outflowEqs}

		Similar to CS94, we follow the spirit of WD67 model to study outflow dynamics. In this model, the poloidal magnetic field is pre-assumed and remains fixed, implying that the magnetic flux surfaces are static. Consequently, each magnetic flux surface is mechanically independent and treated separately. The outflow problem is thus reduced to solving a series of Bernoulli functions on these surfaces, which describe energy conservation along the magnetic field lines.

		However, a complete outflow model requires additional consideration of the self-consistent adjustment of the magnetic field configuration in the transverse direction, beyond the Bernoulli function, to ensure mechanical equilibrium across the magnetic flux surfaces. This adjustment is governed by the Grad-Shafranov equation (also known as the trans-field equation), a second-order nonlinear partial differential equation that describes the magnetic flux function or the poloidal field configuration (e.g., \citealt{BP82, Sakurai85, Sakurai87, Nitta1991, BeskinParev1993, BeskinNokhrina2006, HPY2019}).

		Nevertheless, the coupled solution of the Bernoulli equation and the trans-field equation is mathematically complex and beyond the scope of this study. Following the WD67 model, we first prescribe the poloidal magnetic field and then investigate the acceleration process of outflow generation by solving the relativistic form of the Bernoulli equation.

\subsection{The magnetic field configuraton} \label{subsec: MF}

In this paper, we adopt a simple magnetic field configuration threading the accretion disk proposed by CS94. The magnetic flux (stream function) is given by 

\begin{equation}
	\Phi=B_0 R_0^2 \left\{\left[\left(\frac{R}{R_{0}}\right)^{2}+(1+\zeta)^{2}\right]^{1/2}-(1+\zeta)\right\},
\end{equation}

\noindent in which

\begin{equation}
	\zeta=\frac Z {R_{0}}\tanh\left(\frac Z H\right),
\end{equation}

\noindent where $R$, $Z$ are cylindrical coordinates. $B_0$ denotes the magnitude of the magnetic field and $R_0$ the geometric scale, respectively. $H$ denotes the disk height. As in CS94, we do not intend to explore the disk structure in detail, and $H/R \equiv 0.1$ is used throughout the following. The field components are given by

\begin{equation}
    B_Z = \frac{1}{R}\frac{\partial\Phi}{\partial R},
\end{equation}

\begin{equation}
    B_R = -\frac{1}{R}\frac{\partial\Phi}{\partial Z}.
\end{equation}

\begin{figure}[ht!]
	\plotone{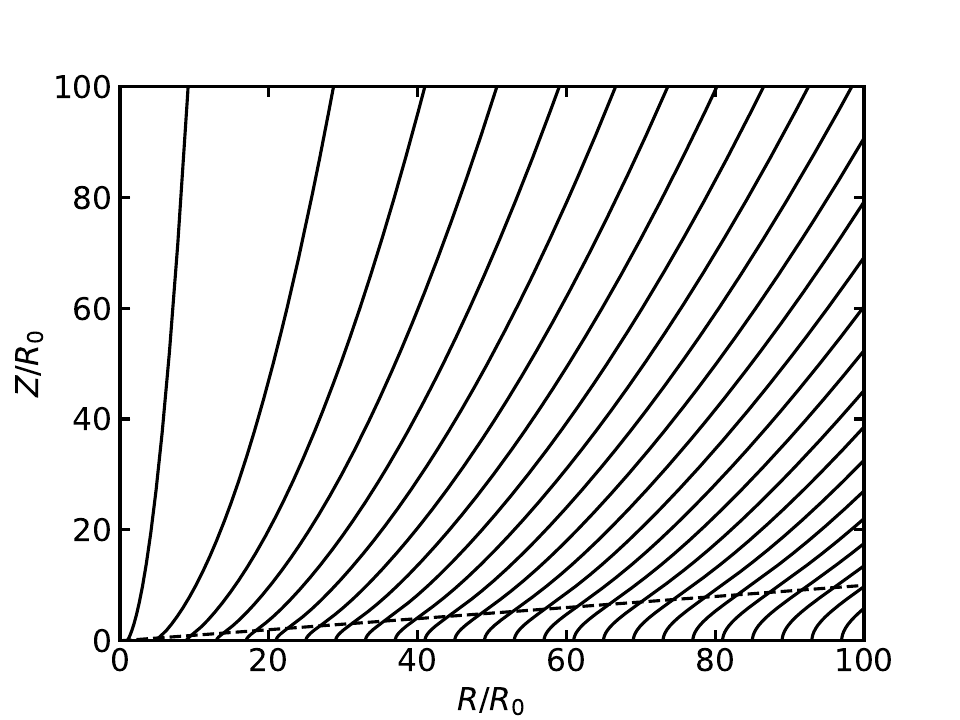}
	\caption{Field configuration of a disk with half disk thickness $H/R=0.1$, with the disk surface denoted by the dashed line. Replot of the Figure 2 in \cite{CaoSpruit1994}.}
	\label{fig:B_field}
\end{figure}

\noindent The magnetical field configuration is shown in Figure \ref{fig:B_field}.

\subsection{Basic equations of the outflow} \label{subsec: BasicEqs}
The laws of mass and momentum-energy conservation, together with Maxwell equations in general relativity, are (\citealt{Landau&Lifshitz1971, Weinberg1972, MTW2018, Anile1989}) 

\begin{equation}\label{eq:continuityEq}
	\nabla_\nu(\rho u^\nu) = 0,
\end{equation}

\begin{equation}\label{eq:energyConserveEq}
	\nabla_\nu(T^{\mu\nu}) = 0,
\end{equation}

\begin{equation}\label{eq:BiancciEq}
	\partial_\mu F_{\nu\lambda} + \partial_\nu F_{\lambda\mu} + \partial_\lambda F_{\mu\nu} = 0,
\end{equation}

\begin{equation}\label{eq:MaxwellEq}
	\nabla_\mu(F^{\mu\nu}) = -J^\nu,
\end{equation}

here $\rho$ is the proper rest-mass density, $u^\nu$ and $J^\nu$ are the 4-velocity and the 4-current density, and $F^\mathrm{\mu\nu}$ is the Faraday tensor of the electromagnetic feld. The stress-energy-momentum tensor is

\begin{equation}\label{eq:stressTensor}
\begin{aligned}
	T^{\mu\nu}&=\rho hu^\mu u^\nu/c^2+pg^{\mu\nu} \\ 
	&+\frac{1}{4\pi}\left(F^{\nu\sigma}F_\sigma^\mu-\frac{g^{\mu\nu}F^{\lambda\kappa}F_{\lambda\kappa}}{4}\right),
\end{aligned}
\end{equation}

\noindent where $g^{\mu\nu}$ is the metric, $p$ is the gas pressure. We assume here a polytropic relation for the ideal gas, i.e., $p / \rho^\Gamma = {K}$, here ${K}$ is a constant, and $\Gamma$ is the polytropic index, so the specific enthalpy is given by
\begin{equation}\label{eq:enthalpy}
	h = c^2 + \frac{\Gamma}{\Gamma-1}\frac{p}{\rho} = c^2 + \frac{\Gamma}{{\Gamma}-1}{K} \rho^{\Gamma-1}.
\end{equation}

In order to close the system, the current density $J_\nu$ must be specified in terms of the other known quantities, through an additional equation, Ohm’s law. In the MHD approximation, Ohm’s law becomes the condition that the net electric field in the fluid frame must vanish, which in covariant form reads

\begin{equation}
	F_{\mu\nu}u^{\nu}=0.
\end{equation}

With this approximation, equations \eqref{eq:continuityEq}-\eqref{eq:MaxwellEq} can be rewritten in term of proper density, pressure, 4-velocity and magnetic field, reducing
to a system of eight equations for eight variables, plus the solenoidal condition on the magnetic field, $\nabla\cdot \boldsymbol{B}=0$.

For simplicity, we don't consider the rotation of the black hole, regardless of its influence of the acceleration of the outflow. Therefore, we employ the Schwarzschild instead of the Kerr metric. In Boyer-Lindquist coordinates $(t, r, \theta, \phi)$,  the metric is given by

\begin{equation}
	\begin{aligned}
	ds^2 &= g_{\mu\nu}dx^\mu dx^\nu \\
	     &= -\alpha^2 dt^2 + \alpha^{-2} dr^2 + r^2 d\theta^2 + r^2\sin^2\theta d\phi^2,
	\end{aligned}
\end{equation}

\noindent in which $\alpha^2 = 1-2G{M_*}/r$, here ${M_*}$ is the mass of the central object, such as either a black hole(BH) or a neutron star(NS).

Given the axisymmetry and steady state, equations \eqref{eq:continuityEq}-\eqref{eq:MaxwellEq} result in six conserved quantities along stream lines (flux tubes) (\citealp{Camenzind1986, Camenzind1986b, Camenzind1987, DaigneDrenkhahn2002, Bucciantini2006}):

\begin{equation}\label{eq:massFlux}
	\mathcal{F}=\alpha\rho\gamma v_{\mathrm{p}}A,
\end{equation}

\begin{equation}\label{eq:magneticFlux}
	\Phi=B_{\mathrm{p}}A,
\end{equation}

\begin{equation}\label{eq:OmegaB1}
	\Omega=\frac{\alpha(v_\phi-B_\phi/B_\mathrm{p}v_\mathrm{p})}{R},
\end{equation}

\begin{equation}\label{eq:L1}
	{L}=R\left(\frac{{h}\gamma v_\phi}{c^2}-\frac{\alpha\Phi B_\phi}{4\pi \mathcal{F}}\right),
\end{equation}

\begin{equation}\label{eq:Etot1}
	\mathcal{E}=\alpha\left({h}\gamma -\frac{\Omega\Phi RB_\phi}{4\pi \mathcal{F}}\right),
\end{equation}

\begin{equation}\label{eq:S}
	{K}=\frac{p}{\rho^\Gamma},
\end{equation}

\noindent where subscript $\mathrm{p}$ and $\phi$ denotes idal and azimuthal components, respectively,  $A$ is the area of the flux tube, $R = r \sin\theta$ the cylindrical radius, and $\gamma$ is the Lorentz factor, i.e.,

\begin{equation}\label{eq:gamma0}
	\gamma^2 = \frac{1}{1-v^2/c^2}.
\end{equation}


From the ratio of equations \eqref{eq:massFlux} and \eqref{eq:magneticFlux}, one gets

\begin{equation}\label{eq:massLoading}
	\frac{\alpha\rho\gamma v_\mathrm{p}}{B_\mathrm{p}}=\kappa, 
\end{equation}

\noindent where $\kappa\equiv \mathcal{F}/\Phi$. And we rewrite \eqref{eq:OmegaB1}, \eqref{eq:L1} and \eqref{eq:Etot1} as follow,

\begin{equation}\label{eq:OmegaB2}
	(\alpha v_\phi-R\Omega)B_\mathrm{p}=\alpha v_\mathrm{p} B_\phi,
\end{equation}

\begin{equation}\label{eq:L2}
	R\left(\frac{{h}\gamma v_\phi}{c^2}-\frac{\alpha B_\phi}{4\pi\kappa}\right)={L},
\end{equation}

\begin{equation}\label{eq:Etot2}
	\alpha\left({h}\gamma -\frac{R\Omega B_\phi}{4\pi\kappa}\right)=\mathcal{E}.
\end{equation}

It is convenient to write the total energy as $\mathcal{E}=c^2+E+L\Omega$ so one can get a simpler form of the energy conservation by subtracting $\Omega\times$\eqref{eq:L2} from equation \eqref{eq:Etot2}, that is

\begin{equation}\label{eq:E}
	{h}\gamma(\alpha -R\Omega v_\phi/c^2)=E+c^2.
\end{equation}

\noindent The poloidal magnetic field $B_\mathrm{p}$ is given in Section \ref{subsec: MF}, and the remaining five variables, $v_\mathrm{p}$, $v_\phi$, $B_\phi$, $\rho$, $h$, can then be solved from the set of equations comprising \eqref{eq:massLoading} through \eqref{eq:L2}, together with \eqref{eq:E} and \eqref{eq:enthalpy}, assuming that the constants $K$, $\Omega$, $\color{red}{L}$, $E$ and $\kappa$ are predetermined.

\subsection{A Bernoulli-like formulation} \label{subsec: BernoulliEq}

Before solving the equations, we would like to clarify two particular positions for each field line, i.e., the Alfv\'en point and the {\it Light Cylinder} (LC) radius.
Combining equations \eqref{eq:massLoading}-\eqref{eq:L2} and \eqref{eq:E}, one has

\begin{equation}\label{eq:gamma}
	\gamma=\frac{(E+c^2)(M^2-1)+M^2L\Omega}{\alpha {h} (M^2-1+R^2\Omega^2/\alpha^2c^2)},
\end{equation}
\noindent where the Alfv\'enic Mach number is defined as

\begin{equation}\label{MachNumber}
	M^{2}\equiv\frac{4\pi\kappa^{2}h}{\alpha^{2}\rho c^2}.
\end{equation}

\noindent The Alfv\'{e}n point is determined by making the numerator and denominator of equation \eqref{eq:gamma} zero at the same time, i.e.,

\begin{equation}\label{eq:M_A}
	M_\mathrm{A}^{2}=1-\frac{R_\mathrm{A}^{2}\Omega^{2}}{\alpha_\mathrm{A}^{2}c^{2}},
\end{equation}

\begin{equation}\label{eq:LEA}
	\frac{L\Omega}{E+c^{2}}=\frac{1-M_\mathrm{A}^{2}}{M_\mathrm{A}^{2}}=\frac{R_\mathrm{A}^{2}\Omega^{2}}{\alpha_\mathrm{A}^{2}M_\mathrm{A}^{2}c^{2}}.
\end{equation}

The LC radius is defined by

\begin{equation}\label{eq:R_LC}
	-\alpha_\mathrm{LC}^2c^2+R_\mathrm{LC}^2\Omega^2=0,
\end{equation}
which corresponds to the last place where corotation is allowed. One sees immediately from \eqref{eq:M_A} and \eqref{eq:R_LC} that the Alfv\'en point stays always inside the LC surface (because of $M_\mathrm{A} > 0$).

To solve the basic equations above, we focus on the energy conservation equation \eqref{eq:E}, which holds a pivotal position akin to the Bernoulli theorem in the classical hydrodynamics. Specially, we will represent $\gamma$ and $v_\phi$ in terms of $\rho$ by making use of the rest equations, thereby obtaining an equation that is solely about $\rho$. In order to do so, we firstly explore the relationships between $\gamma$ and $v_\phi$ by eliminating other variables $B_\phi$ and $h$. The first relation comes from the definition of $\gamma$ itself. For convenience, \noindent We introduce the rescaled quantities

\begin{equation}
	u_\mathrm{p} = \gamma v_\mathrm{p}, \  u_\mathrm{\phi} = \gamma v_\mathrm{\phi},
\end{equation}

\noindent and equation \eqref{eq:gamma0} could be rewritten as 

\begin{equation}\label{eq:gamma2}
	\gamma^2c^2-u_\phi^2-u_\mathrm{p}^2=c^2.
\end{equation}

The component $u_\mathrm{p}$ can be expressed from \eqref{eq:massLoading} and substituted into \eqref{eq:gamma2} to provide the first relation between $\gamma$
and $u_\phi$, that is

\begin{equation}\label{eq:gamma3}
	\gamma^2c^2-u_\mathrm{\phi}^2-\frac{\kappa^2 B_\mathrm{p}^2}{\alpha^2 \rho^2}=c^2.
\end{equation}

\noindent Considering a formula indicated from \eqref{MachNumber}, 

\begin{equation}\label{eq:M_A2}
	M_\mathrm{A}^{2}=\frac{4\pi\kappa^{2} h_\mathrm{A}}{\alpha_\mathrm{A}^{2}\rho_\mathrm{A} c^2},
\end{equation}

\noindent equation \eqref{eq:gamma3} becomes

\begin{equation}\label{eq:gm_vphi1}
	\gamma^2c^2-u_\mathrm{\phi}^2-\frac{\alpha_\mathrm{A}^2M_\mathrm{A}^2c^2}{\alpha^2h_\mathrm{A}}\frac{\rho_\mathrm{A}}\rho\frac{B_\mathrm{p}^2}{4\pi\rho}=c^2.
\end{equation}

Meanwhile, the second relation between those two components could be achieved by eliminating $B_\phi$ from equations \eqref{eq:OmegaB2} and \eqref{eq:L2}, i.e.,

\begin{equation}\label{eq:L3}
	R\left[\frac{h\gamma v_\phi}{c^2} - \frac{(\alpha v_\phi-R\Omega)B_\mathrm{p}}{4\pi\kappa v_\mathrm{p}} \right] = {L},
\end{equation}

\noindent Noting the parallel relationship between $v_\mathrm{p}$ and $B_\mathrm{p}$ in equation \eqref{eq:massLoading}, it follows that

\begin{equation}\label{eq:L4}
	R\left[\frac{h\gamma v_\phi}{c^2} - \frac{\alpha \rho \gamma(\alpha v_\phi-R\Omega)}{4\pi\kappa^2} \right] = {L},
\end{equation}

\noindent equation \eqref{eq:L4} can be rewritten as 

\begin{equation}\label{eq:L5}
	\frac{\gamma h R}{c^2}\left[\left(1-\frac{\alpha^2\rho h_\mathrm{A}}{\alpha_\mathrm{A}^2M_\mathrm{A}^2\rho_\mathrm{A}h}\right)v_\phi + \frac{\alpha\rho h_\mathrm{A}}{\alpha_\mathrm{A}^2M_\mathrm{A}^2\rho_\mathrm{A}h} R\Omega \right] = {L}.
\end{equation}

\noindent equations \eqref{eq:L5} and \eqref{eq:E} could be separately rewritten as 

\begin{equation}\label{eq:L6}
	\frac{h R}{c^2}\left[\left(1-\frac{\alpha^2\rho h_\mathrm{A}}{\alpha_\mathrm{A}^2M_\mathrm{A}^2\rho_\mathrm{A}h}\right)u_\phi + \frac{\alpha\rho h_\mathrm{A}R\Omega}{\alpha_\mathrm{A}^2M_\mathrm{A}^2\rho_\mathrm{A}h}\gamma \right] = {L}.
\end{equation}

\begin{equation}\label{eq:E2}
	{h}(\gamma \alpha -R\Omega u_\phi/c^2)=E+c^2.
\end{equation}

\noindent Making use of the relation in equation \eqref{eq:LEA}, one can get the first relation between $\gamma$ and $u_\phi$ by subtracting $\frac{R_\mathrm{A}^2\Omega}{\alpha_\mathrm{A}^2M_\mathrm{A}^2 c^2} \times$ equation \eqref{eq:E2} from \eqref{eq:L6}, it gives

\begin{equation}\label{eq:gm_vphi2}
	\begin{aligned}
	& \left(1-\frac{\alpha^2\rho h_A}{\alpha_A^2M_A^2\rho_Ah} + \frac{R_A^2\Omega^2}{\alpha_A^2M_A^2c^2}\right)Ru_\phi+ \\
	& \ \ \ \ \frac\alpha{\alpha_A^2M_A^2}\left[\frac{\rho h_A}{\rho_Ah}\left(\frac R{R_A}\right)^2-1\right]R_A^2\Omega\gamma=0.
	\end{aligned}
\end{equation}

Combining equation \eqref{eq:gm_vphi1} and \eqref{eq:gm_vphi2} leads to the expressions of $u_\phi$ and $\gamma$ in terms of $\rho$ and $R$. Taking this two expressions, the energy conservation equation \eqref{eq:E2} could be formally be written as:

\begin{equation}\label{eq:BernoulliEq}
	H(R, \rho; R_\mathrm{A}, \rho_A, K, \Omega, M_*, R_\mathrm{i}, g, {\Gamma}) = E,
\end{equation}

\noindent for clarity, here we list all the involved parameters, in which $g$ stands for the geometry of the field line rooted at $R_\mathrm{i}$, it tells the positional height and then the local gravitational and magnetic field. We restricted our solution to a given magnetic field, so $g$ is regarded as known and not changed, and we fix ${\Gamma}=4/3$ for the whole paper. Meanwhile, we take $\Omega$ to be the Kepler angular momentum which is known as $\Omega=\sqrt{GM_*/R_\mathrm{i}^3}$. Considering $\Omega$ and $K$ are provided, once $E$ is given, $R_\mathrm{A}$ and $M_\mathrm{A}$ are related to ${L}$ in equation \eqref{eq:LEA}, and then $\rho_\mathrm{A}$ is related to $\kappa$ in equation \eqref{eq:M_A2}, resulting in $\rho_\mathrm{i}$ solved from equation \eqref{eq:BernoulliEq}.  From this perspective, we can assert that the three characteristic variables—namely, $\rho_\mathrm{i}$, $\rho_\mathrm{A}$ and $R_\mathrm{A}$—are equivalent to the set of conservation quantities comprising $E$, ${L}$, and $\kappa$, along with $\Omega$ and $K$. Together, these variables and quantities uniquely determine the solution. Consequently, equation \eqref{eq:BernoulliEq} contains all of the information about the system, it acts just like the classic Bernoulli's law.

\subsection{Normalization of the equations}\label{subec:normlize}
For simplicity, we make a normalization for the system based on the Alfv\'en point, with the dimensionless variables defined as 
\begin{equation}
	\begin{aligned}
	x\equiv&R/R_{\mathrm{A}},\quad {z}\equiv Z/R_{\mathrm{A}},\quad y\equiv \rho/\rho_{\mathrm{A}}, \\
	&x_0 \equiv R_0/R_{\mathrm{A}},\quad y_\mathrm{i} \equiv \rho_\mathrm{i}/\rho_{\mathrm{A}},
	\end{aligned}
\end{equation}

\begin{equation}\label{eq:omega}
		\omega\equiv\frac{R_{\mathrm A}^2\Omega^2}{c^2} = \frac{m}{x_0^2}\left(\frac{R_0}{R_i}\right)^3,
\end{equation}

\begin{equation}\label{eq:beta}
		\beta\equiv\frac{B_\mathrm{Ap}^2}{4\pi\rho_\mathrm{A}c^2} =\frac{y_\mathrm{i}}{\rho_\mathrm{i}^{\prime}}\biggl(\frac{B_\mathrm{Ap}}{B_\mathrm{pi}}\biggr)^{2},
\end{equation}

\noindent where the central object mass has been denoted by 

\begin{equation}
	m \equiv R_\mathrm{g}/R_0,
\end{equation}

\noindent here $R_\mathrm{g} \equiv \sqrt{G {M_*}/c^2}$ is the gravitational radius. Throughout this paper, we take $R_0$ as radius of the innermost stable circular orbit (ISCO) of Schwarzschild black hole, specifically $R_0=6R_\mathrm{g}$, and consequently we have $m=1/6$. $\rho_\mathrm{i}^{\prime}$ denotes the dimensionless initial outflow density, i.e., 

\begin{equation}\label{eq:rho_ip}
	\rho_\mathrm{i}^{\prime} \equiv \frac{4\pi\rho_\mathrm{i}c^2}{B_\mathrm{pi}^2}.
\end{equation}

%

\noindent The two relations of $\gamma$ and $u_\phi$ in equations \eqref{eq:gm_vphi1} and \eqref{eq:gm_vphi2} separately have their normalized form as 

\begin{equation}\label{eq:gm_vphi1b}
	\gamma^{2}-\frac{u_{\phi}^{2}}{c^{2}}-\frac{\alpha_{\mathrm A}^{2}M_{\mathrm A}^{2}}{\alpha^{2}\tilde{h}_{\mathrm A}}\frac{\beta}{y^{2}}\biggl(\frac{B_{\mathrm p}}{B_{\mathrm Ap}}\biggr)^{2}=1,
\end{equation}

\begin{equation}\label{eq:gm_vphi2b}
	\begin{aligned}
	&\left(1-\frac{\alpha^2\tilde{h}_\mathrm{A}}{\alpha_\mathrm{A}^2M_\mathrm{A}^2\tilde{h}}y+\frac\omega{\alpha_\mathrm{A}^2M_\mathrm{A}^2}\right)x\frac{u_\phi}c+ \\
	&\quad\quad\quad\quad \frac\alpha{\alpha_\mathrm{A}^2M_\mathrm{A}^2}\biggl[\frac{{\tilde h}_\mathrm{A}}{\tilde h}yx^2-1\biggr]\sqrt{\omega}\gamma=0,
	\end{aligned}
\end{equation}

\noindent in which the gravitational field coefficient $\alpha$ at any position and the Mach number at Alfv\'en point $M_\mathrm{A}$ both take their normalized expressions as

\begin{equation}
	\alpha^2 = 1 - \frac{2mx_0}{\sqrt{x^2+\tilde{z}^2 x_0^2}},
\end{equation}

\begin{equation}
	M_\mathrm{A}^2 = 1 - \omega/\alpha_\mathrm{A}^2,
\end{equation}

\noindent Here $\tilde z\equiv Z/R_0$. Meanwhile, $\tilde{h}\equiv h/c^2$ is presented as 

\begin{equation} 
	\tilde{h} = 1 + \frac{T_\mathrm{i}^\prime}{{\Gamma}-1}\left(\frac{y}{y_\mathrm{i}}\right)^{{\Gamma}-1},
\end{equation}

\noindent here $T_\mathrm{i}^\prime$ denotes the initial thermal sonic speed normalized by light speed, i.e.,

\begin{equation}\label{eq:T_ip}
	T_\mathrm{i}^\prime \equiv \frac{c_\mathrm{s, i}^2}{c^2} \equiv \frac{\Gamma {K} \rho_\mathrm{i}^{\Gamma-1}}{c^2}.
\end{equation}

\noindent The Bernoulli-like equation \eqref{eq:BernoulliEq} is normalized as 

\begin{equation}\label{eq:H2}
	\begin{aligned}
		\tilde{H}(x,y) + 1 &= \tilde{h}(\alpha\gamma-\sqrt{\omega}xu_\phi/c) \\
		&=\tilde{E} + 1,
	\end{aligned}
\end{equation}
\noindent here $\tilde{H}\equiv H/c^2$, $\tilde{E}\equiv E/c^2$. 

 We do not go further to elaborate the mathematically deducing details which lead to the final expressions:

\begin{equation}
	\alpha\gamma=\sqrt{\mathcal{K}(x,y)}\frac{N(x,y)}{\sqrt{\mathcal{D}(x,y)}},
\end{equation}

\begin{equation}
	\frac{xu_\phi}c=\sqrt{\mathcal{K}(x,y)}\frac{D(x,y)}{\sqrt{\mathcal{D}(x,y)}},
\end{equation}

\begin{equation}
	\begin{aligned}
	\tilde{H}(x,y) & =\tilde{h}\sqrt{\mathcal{K}(x,y)}\frac{|\mathcal{N}(x,y)|}{\sqrt{\mathcal{D}(x,y)}}-1\\
	& = \tilde{E},
	\end{aligned}
\end{equation}

\noindent where we have introduced the following auxiliary functions,

\begin{equation}
	N(x,y)=-\alpha^2\tilde{h}_\mathrm{A}y+\alpha_\mathrm{A}^2\tilde{h},
\end{equation}

\begin{equation}
	D(x,y)=\sqrt{\omega}(\tilde{h}-\tilde{h}_\mathrm{A}x^2y),
\end{equation}

\begin{equation}
	\mathcal{N}(x,y)=N(x,y)-\sqrt{\omega}D(x,y),
\end{equation}

\begin{equation}
	\mathcal{D}(x,y)=x^2N^2(x,y)-\alpha^2D^2(x,y),
\end{equation}

\begin{equation}
	\mathcal{K}(x,y)=\alpha^2x^2\left[1+\frac{\alpha_{\mathrm A}^2M_{\mathrm A}^2}{\alpha^2\tilde{h}_{\mathrm A}}\frac{\beta}{y^2}\left(\frac{B_{\mathrm p}}{B_\mathrm{Ap}}\right)^2\right].
\end{equation}

It worthwhile to mention that we have timely hided the parameters in the above dimensional formulas. Taking Bernoulli equation as an example, it has a full formulation as follow,

\begin{equation}
 \tilde{H}(x,y; x_0, y_\mathrm{i}, \rho_\mathrm{i}^\prime, T_\mathrm{i}^\prime, \mathcal{R}_\mathrm{i}, m, g, {\Gamma}) = \tilde{E},
\end{equation}

\noindent here $\mathcal{R}_\mathrm{i}\equiv {R}_\mathrm{i}/{R}_\mathrm{0}$ denotes the footpoint position, $\rho_\mathrm{i}^\prime$ and $T_\mathrm{i}^\prime$ are related to $\rho_\mathrm{i}$ and $K$. Together with $m$, $g$, ${\Gamma}$, these quantities are set as prescribed parameters in the model. And then $x_0$ and $y_\mathrm{i}$ are separately related to $R_A$ and $\rho_A$.

Similar to \cite{DaigneDrenkhahn2002}, it is important to emphasize that the Bernoulli-like function $\tilde{H}(x,y)$ is not defined everywhere in the region where $x>0$ and $y>0$, in contrast to the classical limit, as exemplified in \cite{Sakurai85}. The permitted domain is strictly constrained by the condition that the function $\mathcal{D}(x,y)$ must be positive, which is depicted in Appendix \ref{sec:domainXY}.

\subsection{Properties of the solutions\label{subsec:transonicSol}}
As pointed out in the last paragraph in section \ref{subsec: BernoulliEq}, taking $\Omega$ as the Kepler angular velocity, the motion of the matter is determined by four quantities: $R_A$, $\rho_A$, $\rho_\mathrm{i}$ and $K$. In the model, $\rho_\mathrm{i}$ and $K$ are prescribed parameters denoted as $\rho_\mathrm{i}^\prime$ and $T_\mathrm{i}^\prime$. This leaves $R_A$, $\rho_A$ as adjustable quantities, corresponding to different solutions shown as contours of function $H$ in $R-\rho$ plane. From the perspective of initial condition, once $\Omega$ and $K$ are set, the initial values of $\rho_\mathrm{i}$ (initial density), $v_\mathrm{p, i}$ (initial poloidal velocity), and $v_{\phi, \mathrm i}$ (initial azimuthal velocity) are uniquely equivalent to the constants $\kappa$, ${L}$ and $E$, thereby $R_A$ and $\rho_A$. In principle, the initial condition of the matter could be any case. However, as described by \cite{Sakurai85} and \cite{DaigneDrenkhahn2002}, an outflow solution that extends to infinity must pass two saddle points, the slow (S) and fast (F) critical points, here the function $H$ is locally flat in the $R$-$\rho$ plane, i.e.,

\begin{equation}
	\left.\frac{\partial H}{\partial R}\right|_\mathrm{s, f} = \left.\frac{\partial H}{\partial\rho}\right|_\mathrm{s, f}=0,
    \label{eq: saddleEQ}
\end{equation}

\begin{figure}[t!h!]
	\plotone{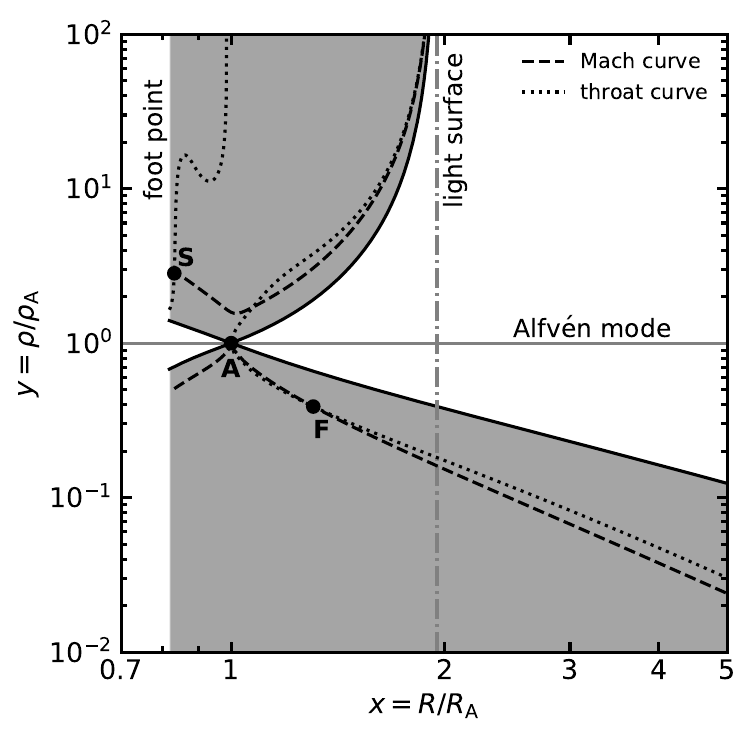}
	\caption{Solution plane of the outflow equations. The gray region corresponds to the domain where the Bernoulli function is well defined. In the sub-Alfv\'enic region ($y>1$), it is limited by the light surface. The gray solid horizontal line (Alfv\'en mode) separates the sub- and super-Alfv\'enic modes. The dashed line indicates the slow ($y>1$) and fast ($y<1$) mode Mach curves and the dotted line the gravitational throat curve. The slow (S) and fast (F) critical points are the intersections of the Mach and throat curves. The Alfv\'en point is denoted by A. This calculation has been made for $R_\mathrm{i} = R_0$ and the parameters ($\rho_\mathrm{i}^\prime = 0.01$, $T_\mathrm{i}^\prime = 50$) have been chosen so that the different points are well separated.
		\label{fig:saddles}}
\end{figure}

\begin{figure}[t!h!]
	\plotone{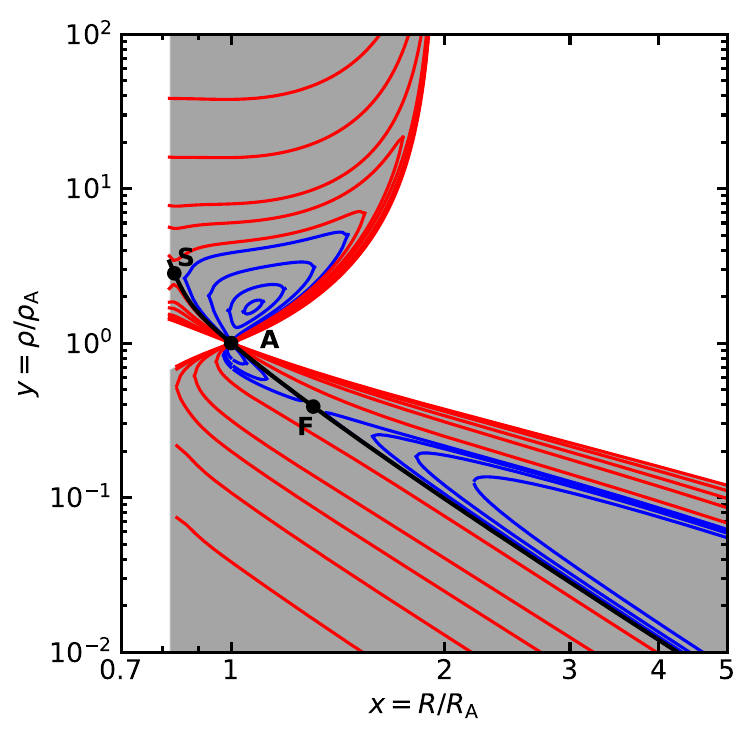}
	\caption{A few level contours of the Bernoulli function on the solution plane described in Figure \ref{fig:saddles}. The outflow solution (thick black line) starts in the sub-Alfv\'en region passes the slow critical point (S) then reaches the Alfv\'en point(A) and enters the super-Alfv\'en region where it passes the fast critical point (F). The red (blue) lines are the level contours with specific energies larger (smaller) than that of the outflow solution.
		\label{fig:contoursH}}
\end{figure}

\noindent Indeed, the two types of curves defined in $R$-$\rho$ plane by $\partial{H}/\partial \rho = 0$ and $\partial{H}/\partial R = 0$ are separately called Mach curve and gravitational throat curve, the S/F point are just the cross point of the slow/fast mode Mach curve and the throat curve. Equation \eqref{eq: saddleEQ} has its dimensionless expression as

\begin{subequations}\label{eq:saddle}
\begin{align}
\frac{\partial \tilde H}{\partial x}(x_\mathrm{s}, y_\mathrm{s}; x_0, y_\mathrm{i})&=\frac{\partial \tilde H}{\partial y}(x_\mathrm{s}, y_\mathrm{s}; x_0, y_\mathrm{i})=0,\\
\frac{\partial \tilde H}{\partial x}(x_\mathrm{f}, y_\mathrm{f}; x_0, y_\mathrm{i})&=\frac{\partial \tilde H}{\partial y}(x_\mathrm{f}, y_\mathrm{f}; x_0, y_\mathrm{i})=0.
\end{align}
\end{subequations}

The fact that the same contour originating from the start point passes through both points $\mathrm S$ and $\mathrm F$ implies

\begin{equation}\label{eq:contour}
	{\tilde H}(x_\mathrm{s}, y_\mathrm{s}; x_0, y_\mathrm{i})={\tilde H}(x_\mathrm{f}, y_\mathrm{f}; x_0, y_\mathrm{i})={\tilde H}(x_\mathrm{i}, y_\mathrm{i}; x_0, y_\mathrm{i}).
\end{equation}

\noindent Given the fixed $\rho_\mathrm{i}$ and $K$---namely $\rho_\mathrm{i}^\prime$ and $T_\mathrm{i}^\prime$---the A point (\{$R_{\mathrm A}$, $\rho_{\mathrm A}$\} or \{$x_\mathrm{0}$, $y_\mathrm{i}$\}) together with the  \text{S} point (\{$R_{\mathrm s}$, $\rho_{\mathrm s}$\} or \{$x_\mathrm{s}$, $y_\mathrm{s}$\}) and the F point (\{$R_{\mathrm f}$, $\rho_{\mathrm f}$\} or \{$x_\mathrm{f}$, $y_\mathrm{f}$\}) could be determined by the 6 individual equations comprising \eqref{eq:saddle} and \eqref{eq:contour}, resulting a unique solution for the motion along a field line from the foot point to infinity. 

Figure \ref{fig:saddles} shows the $x-y$ plane for a particular choice of the parameters with the slow/fast mode Mach curve, the gravitational throat curve. Different level contours of the function $\tilde{H}(x, y)$ are shown in Figure \ref{fig:contoursH}, The solution is one of the contours which passes through the slow and fast points.

\subsection{The constraints for parameters setting}\label{subsec:para}

For convenience, it is useful to express the relation between the  parameters ($\rho_\mathrm{i}^\prime$ and $T_\mathrm{i}^\prime$) and more usual physical quantities. As for the sound speed $c_\mathrm{ss}$, it could have a reference value from the vertical static balance condition in classic disk theory, i.e, $c_\mathrm{ss} \sim H \Omega_\mathrm{K}$, as a result, one has 

\begin{equation}\label{eq:Theta_estimation}
\begin{aligned}
	T_\mathrm{i}^\prime \equiv\left(\frac{c_\mathrm{ss,i}}{c}\right)^2 &\sim \left(\frac{H \Omega_\mathrm{K}}{c}\right)^2 = \left(\frac{H}{R}\right)^2\frac{R_\mathrm{g}}{R} \\
	&= 0.01 \left(\frac{H/R}{0.1}\right)^2\frac{R_\mathrm{g}}{R}.
\end{aligned}
\end{equation}

\begin{figure*}[h!t!]
	\plotone{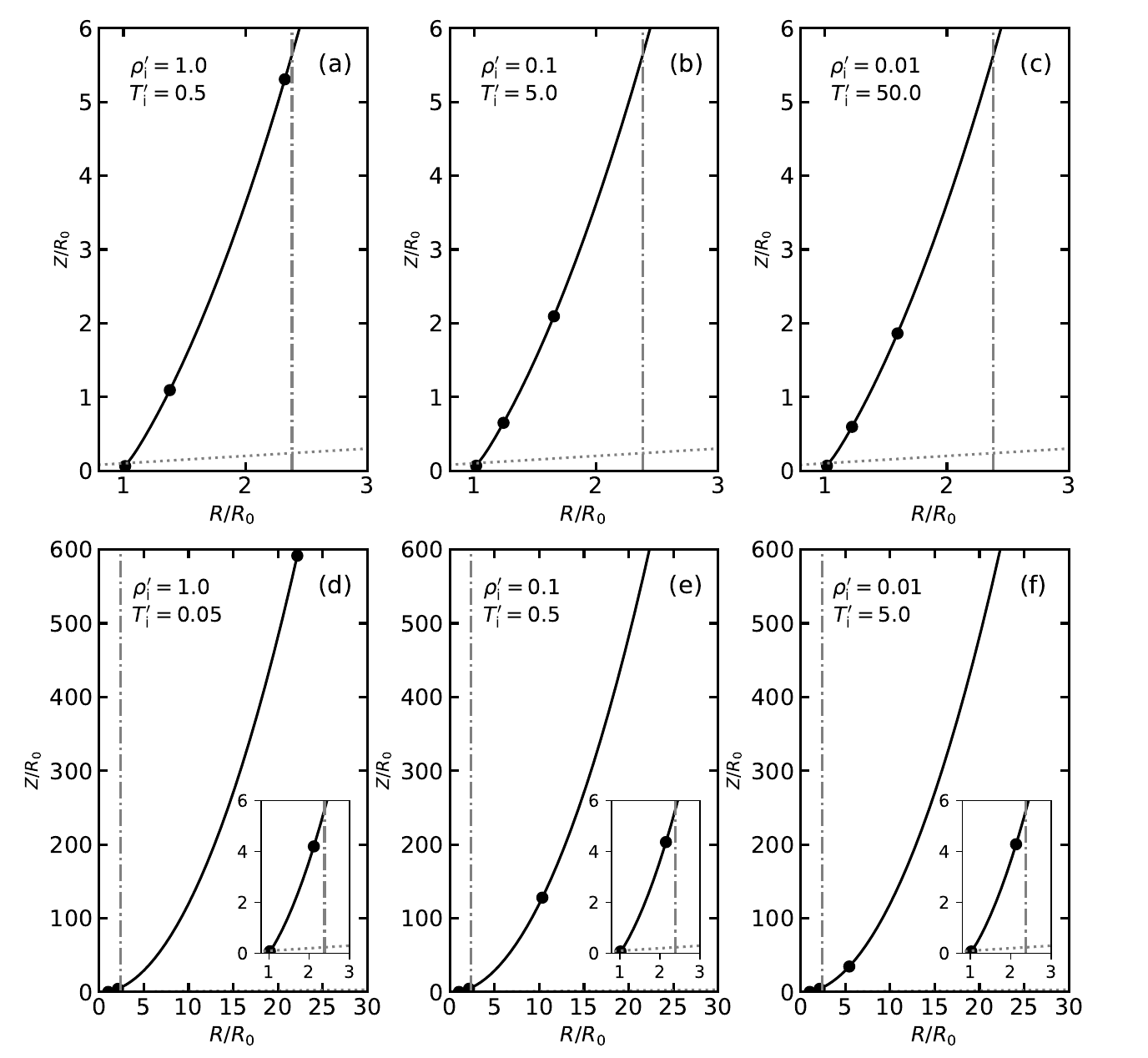}
	\caption{The critical points (slow, Alfv\'en and fast magnetosonic point) along the field line rooted at $R_i/R_0 = 1$. The vertical dashed lines denote light cylinders. Each panel corresponds to different values of the parameter $\rho_{\mathrm i}^\prime$ and $T_{\mathrm i}^\prime$. 
		\label{fig:profiles}}
\end{figure*}

\begin{table}[h!t!]
	\caption{Typical parameters for the field line rooted at $\mathcal{R}_{\mathrm i} = 1$ \label{tab:parasets}}
	\begin{center}
		\begin{tabular}{ccllllcc}
			\hline	
			\multicolumn{2}{l}{case label} & \multicolumn{2}{l}{$\rho_{\mathrm i}^\prime$} & \multicolumn{2}{l}{$T_\mathrm{i}^\prime$} & \multicolumn{2}{c}{$\beta_\mathrm{b}$} \\
			\hline
			\multicolumn{2}{c}{a} & \multicolumn{2}{l}{1.0} & \multicolumn{2}{l}{0.5} & \multicolumn{2}{c}{\multirow{3}{*}{1.0}} \\
			\cline{1-6}\multicolumn{2}{c}{b} & \multicolumn{2}{l}{0.1} & \multicolumn{2}{l}{5.0} & \\
			\cline{1-6}\multicolumn{2}{c}{c} & \multicolumn{2}{l}{0.01} & \multicolumn{2}{l}{50.0} & \\
			\hline
			\multicolumn{2}{c}{d} & \multicolumn{2}{l}{1.0} & \multicolumn{2}{l}{0.05} & \multicolumn{2}{c}{\multirow{3}{*}{0.1}} \\
			\cline{1-6}\multicolumn{2}{c}{e} & \multicolumn{2}{l}{0.1} & \multicolumn{2}{l}{0.5} & \\
			\cline{1-6}\multicolumn{2}{c}{f} & \multicolumn{2}{l}{0.01} & \multicolumn{2}{l}{5.0} & \\
			\hline
		\end{tabular}
	\end{center}
\end{table}

\begin{figure*}[h!t!]
	\gridline{\fig{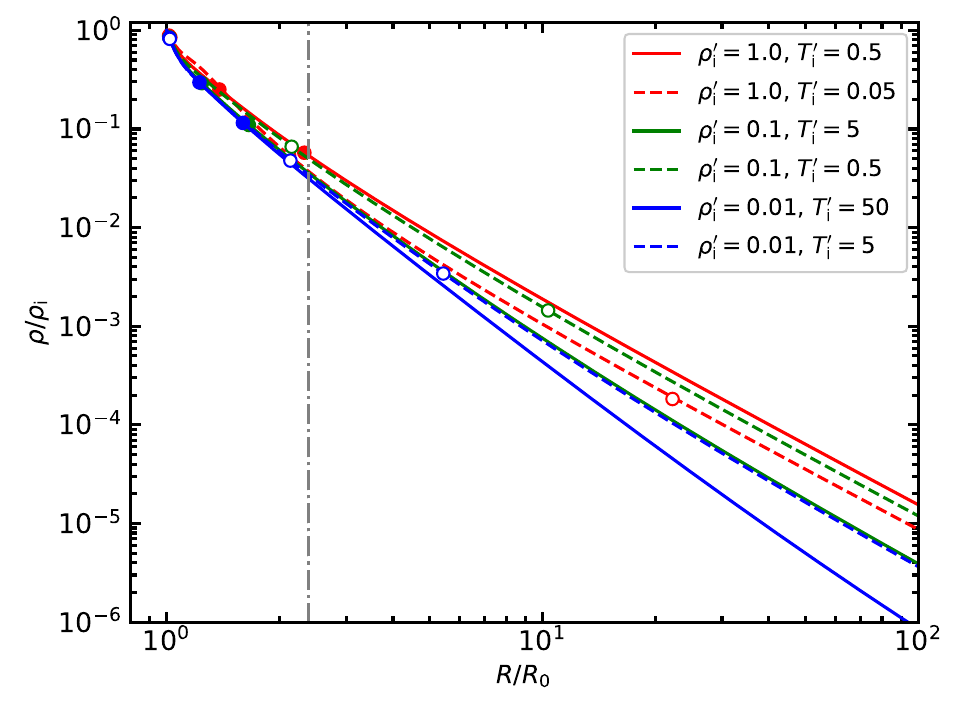}{0.5\textwidth}{(a)}
		\fig{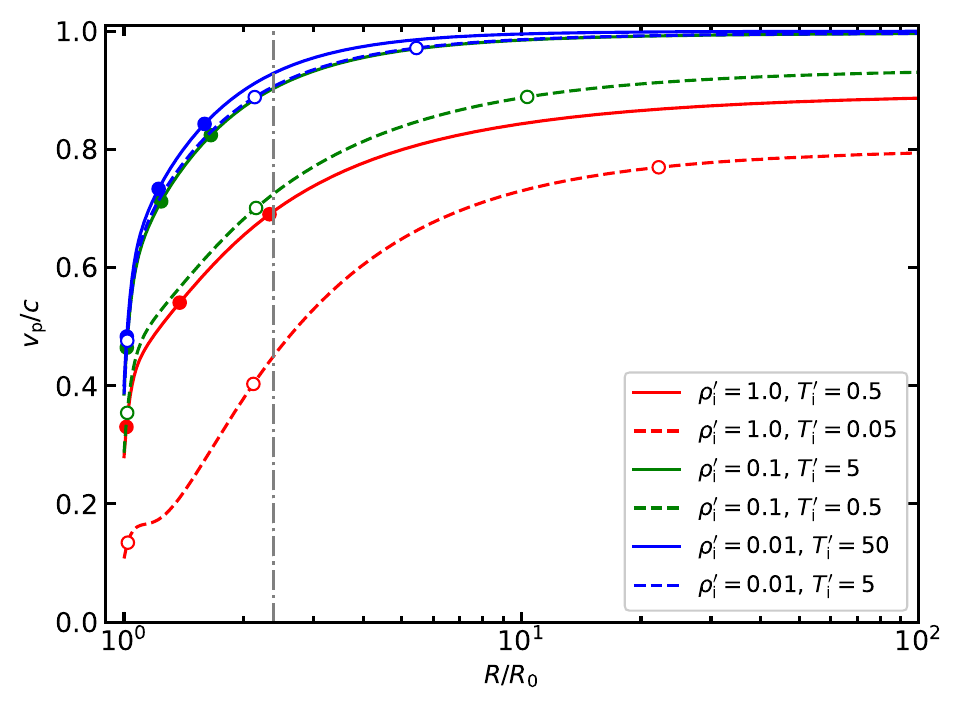}{0.5\textwidth}{(b)}
	}
	\gridline{\fig{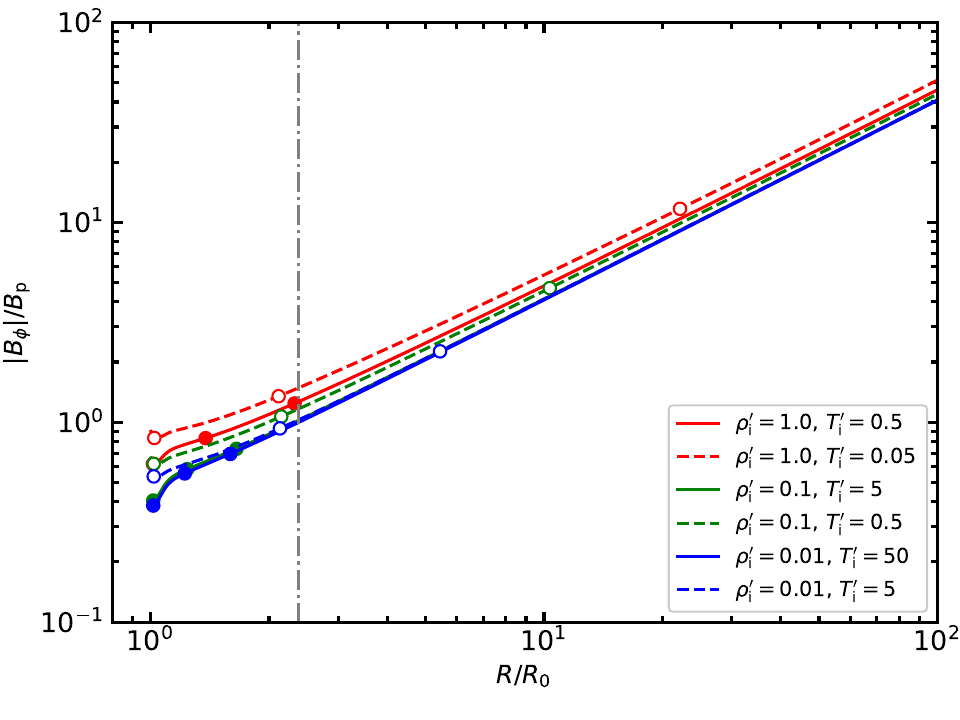}{0.5\textwidth}{(c)}
		\fig{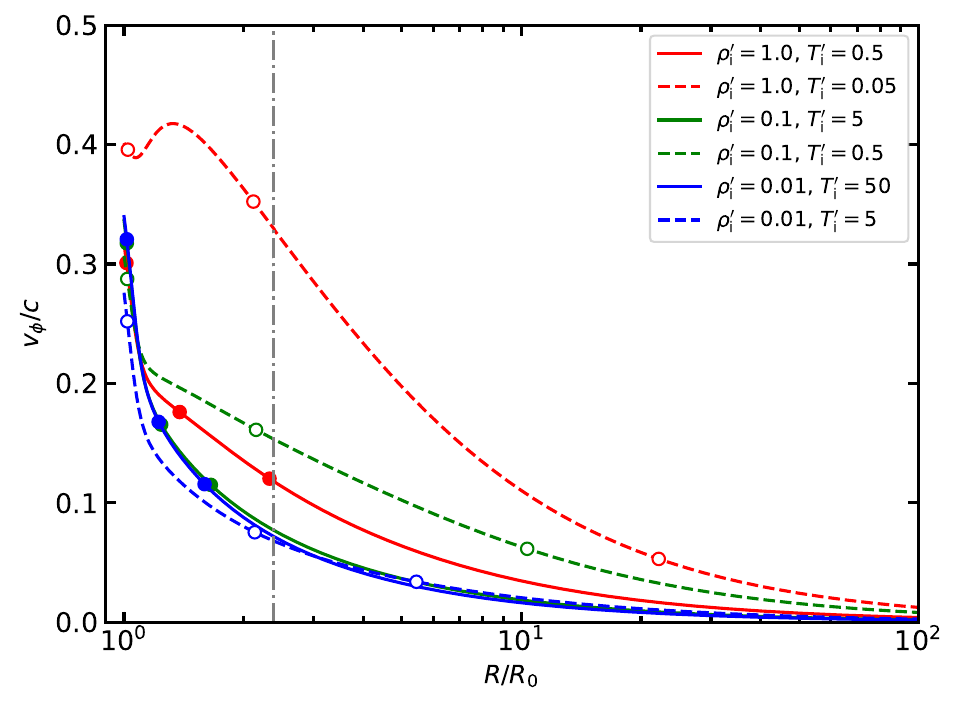}{0.5\textwidth}{(d)}
	}
	\gridline{\fig{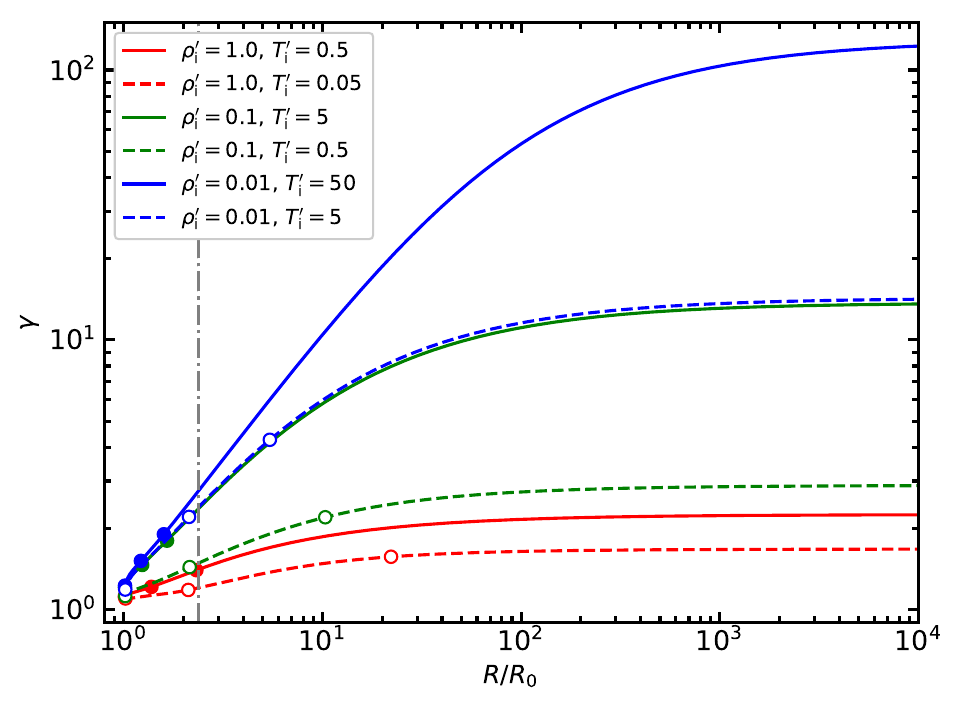}{0.5\textwidth}{(e)}
	}
	\caption{The variables such as density, poloidal velocity, toroidal magnetic field, toroidal velocity and Lorentz factor changing along with the field line rooted at $R_i/R_0 = 1$. The vertical dashed lines denote light cylinders.
		\label{fig:dvbg}}
\end{figure*}

\begin{figure*}[h!t!]
	\plotone{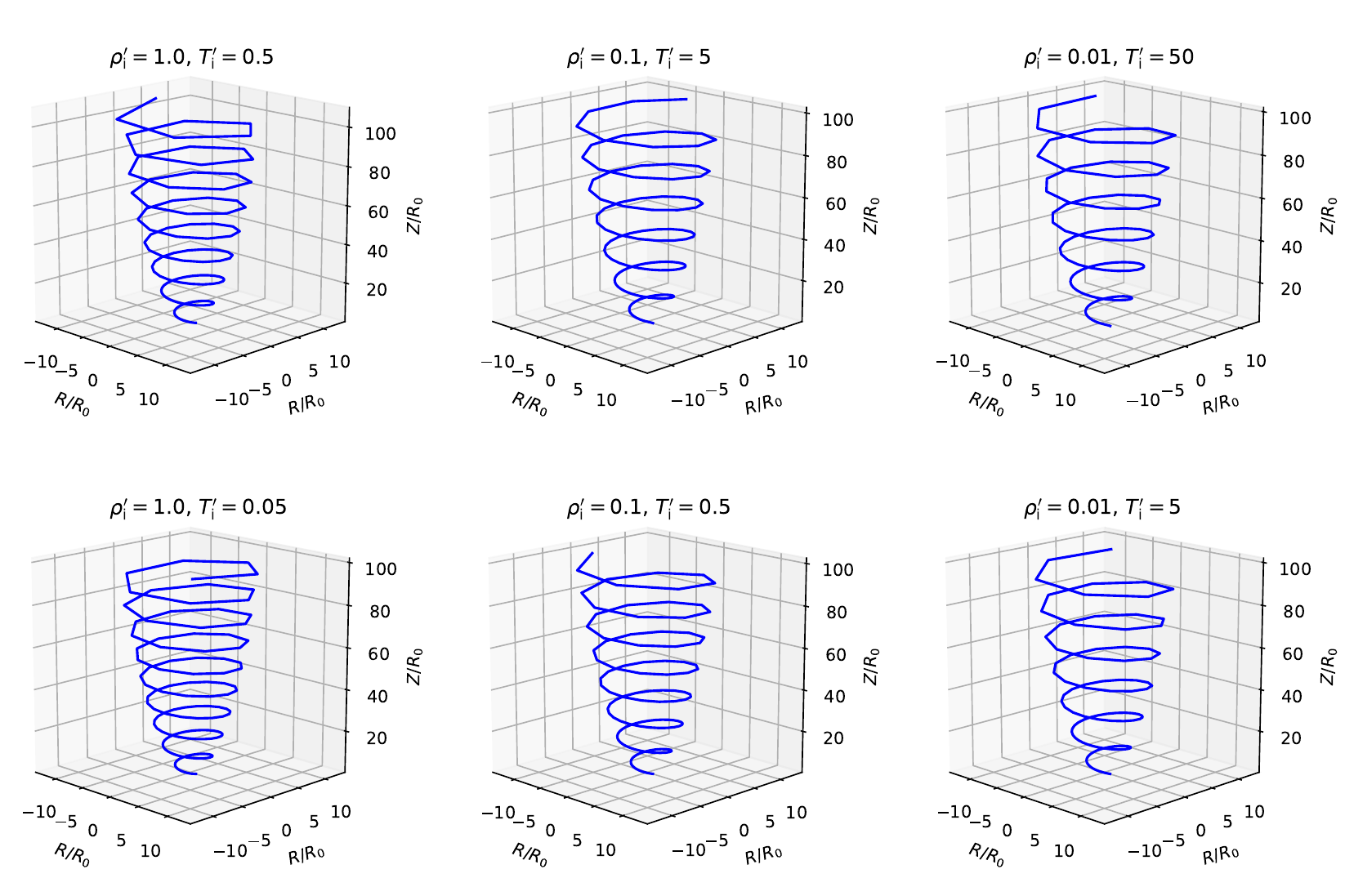}
	\caption{The 3D field line rooted at $R_\mathrm{i}/R_0 = 1$. Each panel corresponds to different values of the parameter $\rho_{\mathrm i}^\prime$ and $T_{\mathrm i}^\prime$. 
		\label{fig:bline3D}}
\end{figure*}

Since the outflow originates from the disk, it is meaningful to estimate the gas density of the disk, which could be related to the accretion disk as follow,

\begin{equation}
	\rho_\mathrm{d} = \frac{\dot{M}}{4\pi R H v_\mathrm{acc}} \sim \frac{\dot{M}}{4\pi R H \alpha c_\mathrm{ss}},
\end{equation} 

\noindent a dimensionless form could be 

\begin{equation}\label{eq:rho_dp}
	\rho_\mathrm{d}^{\prime} \equiv \frac{4\pi\rho_\mathrm{d}c^2}{B_\mathrm{pd}^2}=\frac{\dot{M}c^2}{B_\mathrm{pd}^2 R H \alpha c_\mathrm{ss}}.
\end{equation}

\noindent Considering the magnetic pressure is the same magnitude of ram pressure of the accretted matter, i.e., $B_\mathrm{pd}^2\sim \dot{M}c/4\pi R H$, then one has

\begin{equation}\label{eq:rho_dp2}
	\rho_\mathrm{d}^{\prime} \sim {4\pi }{\alpha}^{-1} (c_\mathrm{ss}/c)^{-1}.
\end{equation}

In addition, combing with equation \eqref{eq:Theta_estimation} it reads

\begin{equation}\label{eq:rho_dp3}
	\begin{aligned}
	\rho_\mathrm{d}^{\prime} &\sim \frac{4\pi}{\alpha}\left(\frac{H}{R}\right)^{-1}\frac{R}{R_\mathrm{g}} \\
	&\sim 1256\left(\frac{\alpha}{0.1}\right)^{-1}\left(\frac{H/R}{0.1}\right)^{-1}\frac{R}{R_\mathrm{g}}.
	\end{aligned}
\end{equation}

Only a little part of the disk matter come into the outflow, and the mass density at the base of the outflow must be much smaller than the disk matter density, i.e., 
\begin{equation}
	\rho_\mathrm{i}^\prime \ll \rho_\mathrm{d}^\prime.
\end{equation}

Furthermore, the magnetic field in our model is strong enough to enforce the gas to corotate with the field line. So the magnetic pressure is assumed
to dominate the gas pressure at the bottom of the outflow, and the reasonable parameters should guarantee the base plasma beta be smaller than unity, i.e.,

\begin{equation}
	\beta_\mathrm{b} = \frac{8\pi \rho_\mathrm{i} c_\mathrm{ss, i}^2}{B_\mathrm{p,i}^2} = 2\rho_{\mathrm i}^\prime T_\mathrm{i}^\prime \lesssim 1.
\end{equation}

\section{Results and Discussions}\label{sec:results}

\subsection{The outflow solution for a certain field line}
In this section, we display the solutions of different parameters with the field line kept unchanged so as to investigate the parameter dependencies of the model. Without loss of generality, we set the footpoint of field line fixed at $R_\mathrm{i}/R_0 = 1$. Aiming to clarify the characteristics of the solutions, we elaborate the results of six groups of parameters, which are listed in Table \ref{tab:parasets} and marked by (a)-(f) separately.

Figure \ref{fig:profiles} shows the three critical points (slow, Alfv\'en and fast magnetosonic point) along the field line under different parameters. One can found the results are sensitive with the two parameter $\rho_\mathrm{i}^\prime$ and $T_\mathrm{i}^\prime$. With the same parameter $\rho_\mathrm{i}^\prime$, as temperature $T_\mathrm{i}^\prime$ rises, these three critical points, on the whole, shift towards the bottom and draw closer to each other. On the contrary, as temperature drops, they generally move farther from the base and spread out more widely in their distribution. Similarly, when parameter $T_\mathrm{i}^\prime$ is fixed, as $\rho_\mathrm{i}^\prime$ increases, the spatial distribution of these three points concentrates towards the bottom.

Figure \ref{fig:dvbg} shows the variables such as density ($\rho/\rho_\mathrm{i}$), poloidal velocity($v_\mathrm{p}/c$), toroidal velocity ($v_\phi/c$), ratio of toroidal to poloidal component of magnetic field ($B_\phi/B_\mathrm{p}$) and Lorentz factor ($\gamma$) changing along with the field line rooted at $R_\mathrm{i}/R_0 = 1$. It is obvious that the poloidal velocity increases accomplied with the decreasing density and the increasing ratio of $B_\phi/B_\mathrm{p}$. This indicates that the gas is accelerated due to magnetic force and thermal expansion. In general, a higher initial temperature $T_\mathrm{i}^\prime$ results in a larger terminal velocity. For most of the six cases, the effect of the magnetic centrifigual accelaration is not significant compared with the magnetic gradient force except for case (d) ($\rho_\mathrm{i}^\prime=1.0$, $T_\mathrm{i}^\prime = 0.05$, red dashed lines in the Figure \ref{fig:dvbg}), for such a case the $v_\phi$ presents a short increase in the vicinity of the bottom which implies a contribution of the magnetically centrifugal force.

Figure \ref{fig:bline3D} shows the 3D field line rooted at $R_\mathrm{i}/R_0 = 1$ under different parameters. Under a fixed $\rho_\mathrm{i}^\prime$, a lower temperature corresponds to a more toroidally tangled magnetic field line. It is understandable if one note that a lower temperature usually implies a larger $v_\phi$ and consequently a larger $B_\phi/B_\mathrm{p}$ as shown in Figure \ref{fig:profiles}.

\begin{figure}[h!t!]
	\plotone{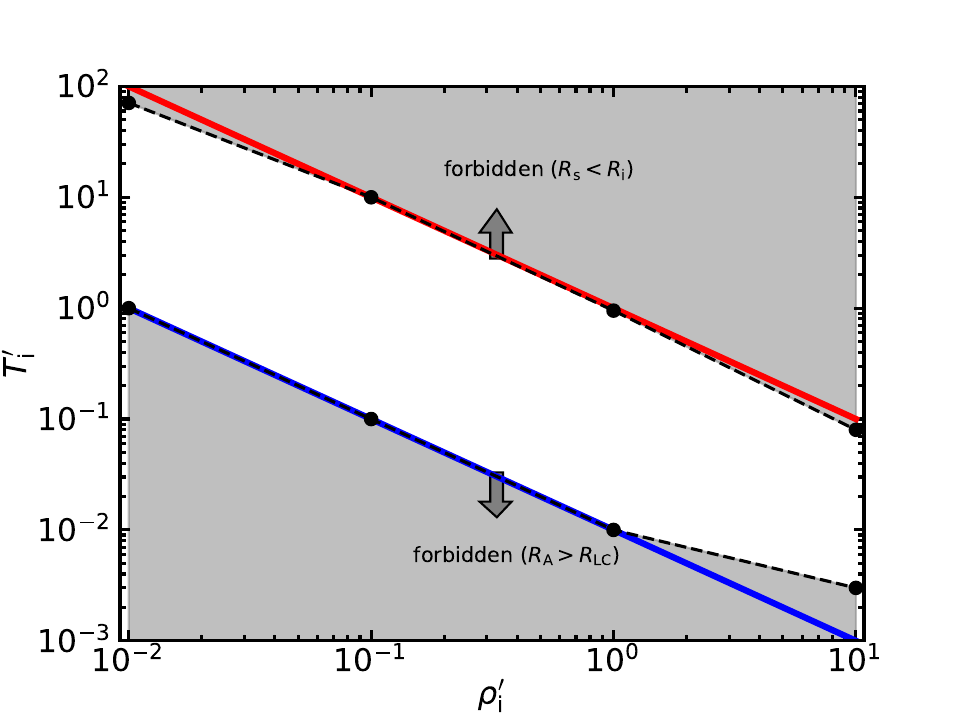}
	\caption{The parameter space for the outflow solutions at $R_\mathrm{i}/R_0 = 1$. 
		\label{fig:dompara}}
\end{figure}

Before exploring the distribution of the outflow along the radius of the disk, it is worthwhile to investigate the parameter space about $\rho_\mathrm{i}^\prime$ and $T_\mathrm{i}^\prime$. As discussed above and shown in Figure \ref{fig:profiles}, for a given $\rho_\mathrm{i}^\prime$, when the $T_\mathrm{i}^\prime$ increases, the corresponding slow point gradually drops down and approaches to the footpoint, this property gives an upper limit for $T_\mathrm{i}^\prime$ under which the slow point coincide with the footpoint. The temperature higher than this upper limit will cause to a slow point at the inner side the footpoint, i.e., $R_\mathrm{s}<R_\mathrm{i}$. This phenomenon is also reported in \cite{CY2020}. On the other side, when the $T_\mathrm{i}^\prime$ decreases, the Alfv\'en point raises up untill it comes to the light cylinder, this gives the lower limit of $T_\mathrm{i}^\prime$. The domain of the parameter space are plotted in Figure \ref{fig:dompara}, in which the permitted and forbidden region are displayed in white and gray color separately. It is found that the upper limit of the parameters approximately complies with a correlation that $\beta_\mathrm{b} = \rho_\mathrm{i}^\prime T_\mathrm{i}^\prime \lesssim 2$ (denoted by red line in Figure \ref{fig:dompara}), while the lower limit approximately obeys $\beta_\mathrm{b} = \rho_\mathrm{i}^\prime T_\mathrm{i}^\prime \gtrsim 0.02$ (denoted by blue line in Figure \ref{fig:dompara}). There is no solution for the parameters located in the forbidden region, in that cases the outflow might be accelerated by any other mechanism, wich is beyond the scope of this paper.

\begin{figure}[h!t!]
	\includegraphics[width=0.4\textwidth]{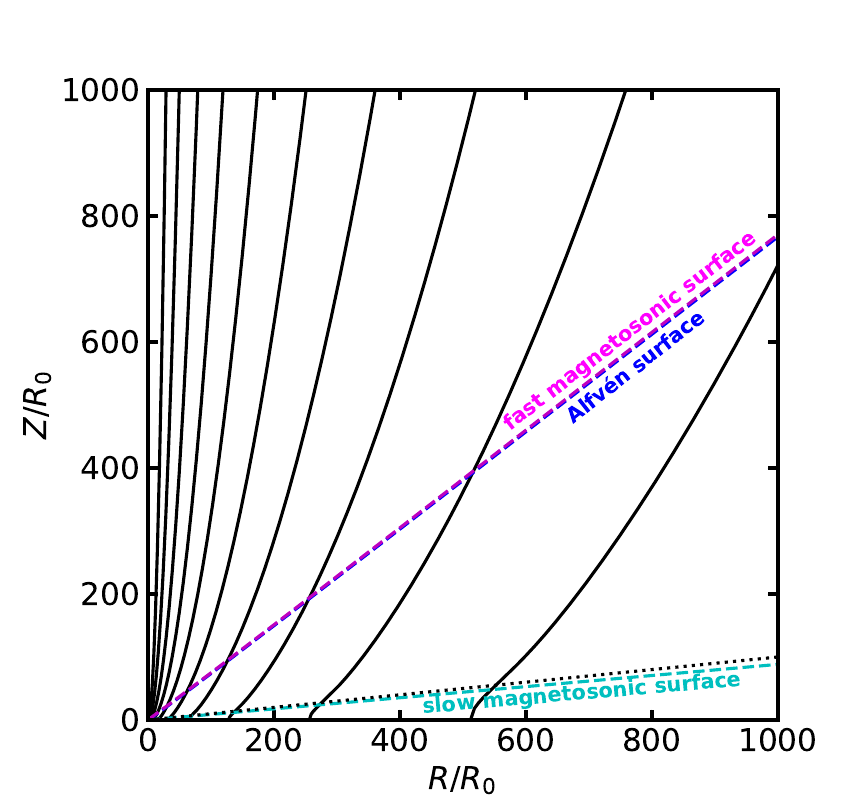}\\
	\includegraphics[width=0.4\textwidth]{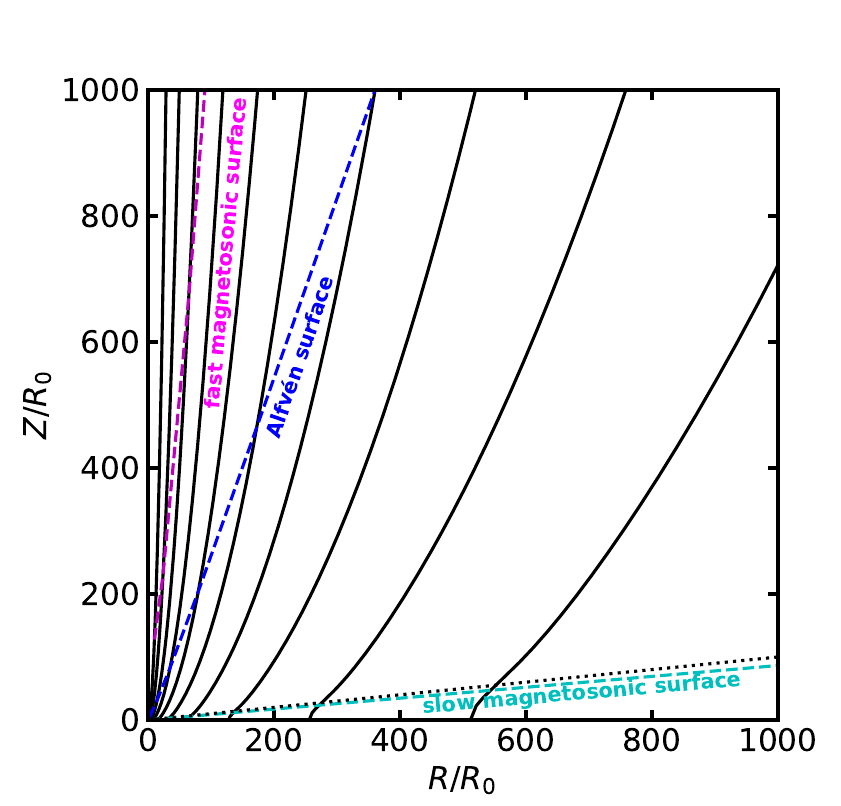}
	\caption{The spatial distribution of the critical points. The upper and lower panel separately corresponds to \{$T_0^\prime = 50$, $\beta_\mathrm{b}=1.0$\} and \{$T_0^\prime = 0.5$, $\beta_\mathrm{b}=0.1$\}. 
		\label{fig:blines}}
\end{figure}

\begin{figure}[h!t!]
    \centering
    \includegraphics[width=0.45\textwidth]{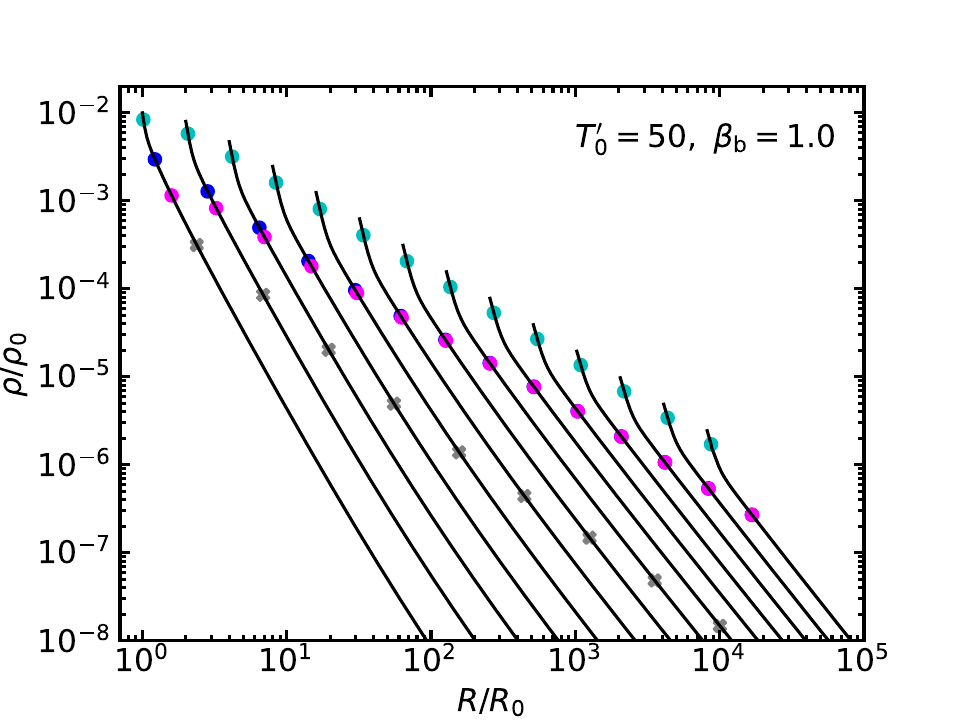}\\
    \includegraphics[width=0.45\textwidth]{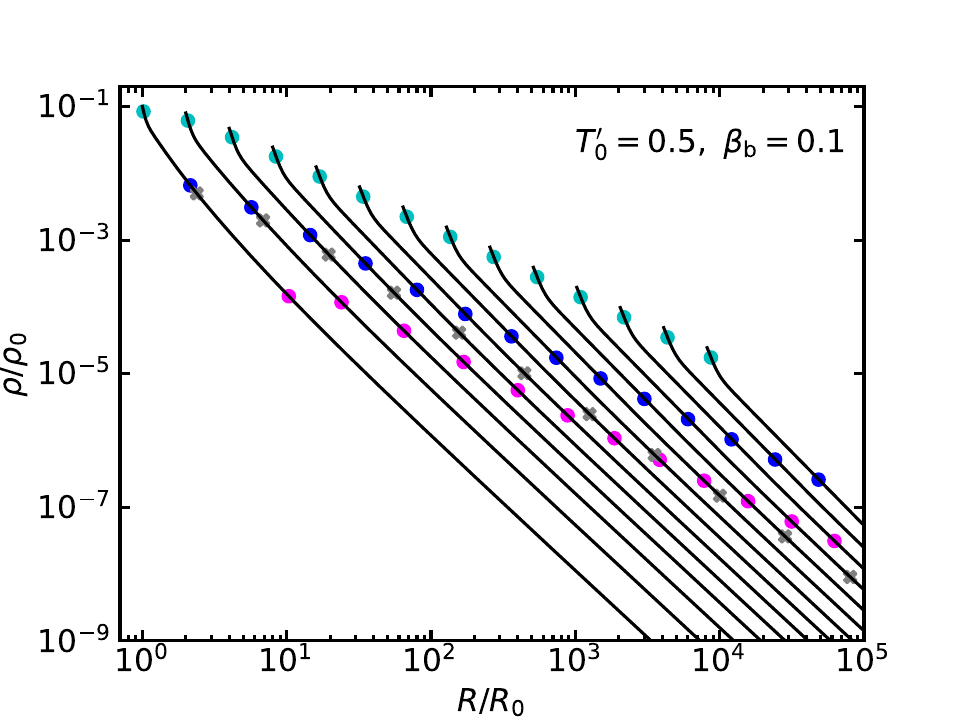}
    \caption{The change of density along a series of magnetic field lines rooted at different disk radii. The upper and lower panel separately corresponds to \{$T_0^\prime = 50$, $\beta_\mathrm{b}=1.0$\} and \{$T_0^\prime = 0.5$, $\beta_\mathrm{b}=0.1$\}. For each line, the slow, Alfv\'en, and fast points are marked by cyan, blue, and magenta circles respectively; the light cylinder is marked by a gray cross.
    \label{fig:Rho_dist}}
\end{figure}

\subsection{The spatial distribution of the outflow \label{subsec: distribution}}

In this section we explore the distribution of the outflow along with disk radius. For doing this, the parameter $T_{\mathrm i}^\prime$ is assumed to be linearly decreasing with the radius in form of $T_{\mathrm i}^\prime = T_0^\prime R_0/R_{\mathrm i}$, as discussed before. Meanwhile, $\rho_\mathrm{i}^\prime$ is adjusted to guarantee plasma beta $\beta_\mathrm{b}$ to be constant, i.e., $\rho_\mathrm{i}^\prime = \beta_\mathrm{b}/2T_{\mathrm i}^\prime$. For comparison, two cases are carefully calculated, one is $T_{0}^\prime = 50$ and $\beta_\mathrm{b}=1.0$, the other is $T_{0}^\prime = 0.5$ and $\beta_\mathrm{b}=0.1$. In addition, for comparision among different field lines, we have normalized the density with a typical density related to the magnetic strength at $R_0$, that is

\begin{equation}
	\rho_0 \equiv \frac{B_\mathrm{p0}^2}{4\pi c^2}
\end{equation}
\noindent the initial density at each foot point can be written as 

\begin{equation}
	\rho_\mathrm{i} = \rho_0 \rho_\mathrm{i}^\prime \left(\frac{B_\mathrm{pi}}{B_\mathrm{p0}}\right)^2
\end{equation}
\noindent and then the density at any position along a field line can be normalized as 

 \begin{equation}
 	\frac{\rho}{\rho_0} = \frac{y}{y_\mathrm{i}} \rho_\mathrm{i}^\prime \left(\frac{B_\mathrm{pi}}{B_\mathrm{p0}}\right)^2
 \end{equation}

\begin{figure*}[h!t!]
	\includegraphics[width=0.5\textwidth]{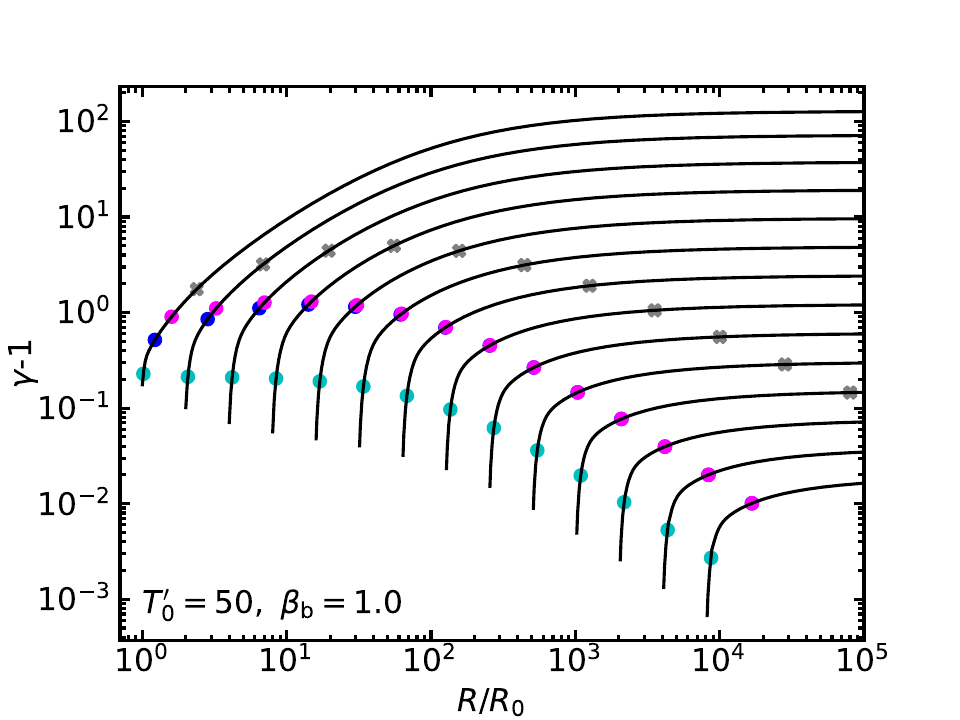}
	\includegraphics[width=0.5\textwidth]{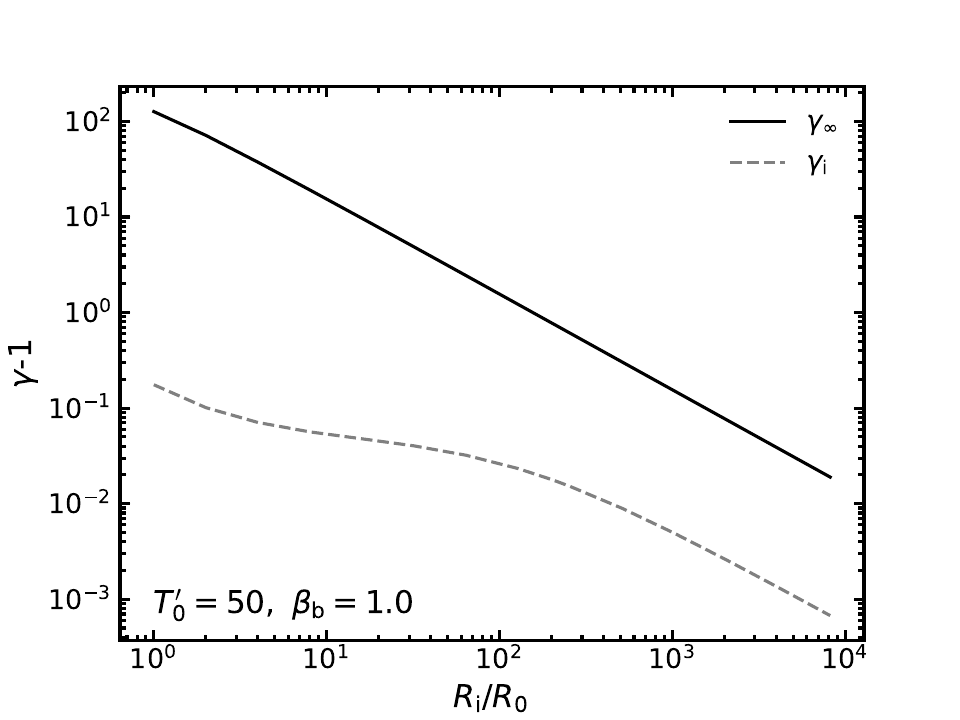}\\
	\includegraphics[width=0.5\textwidth]{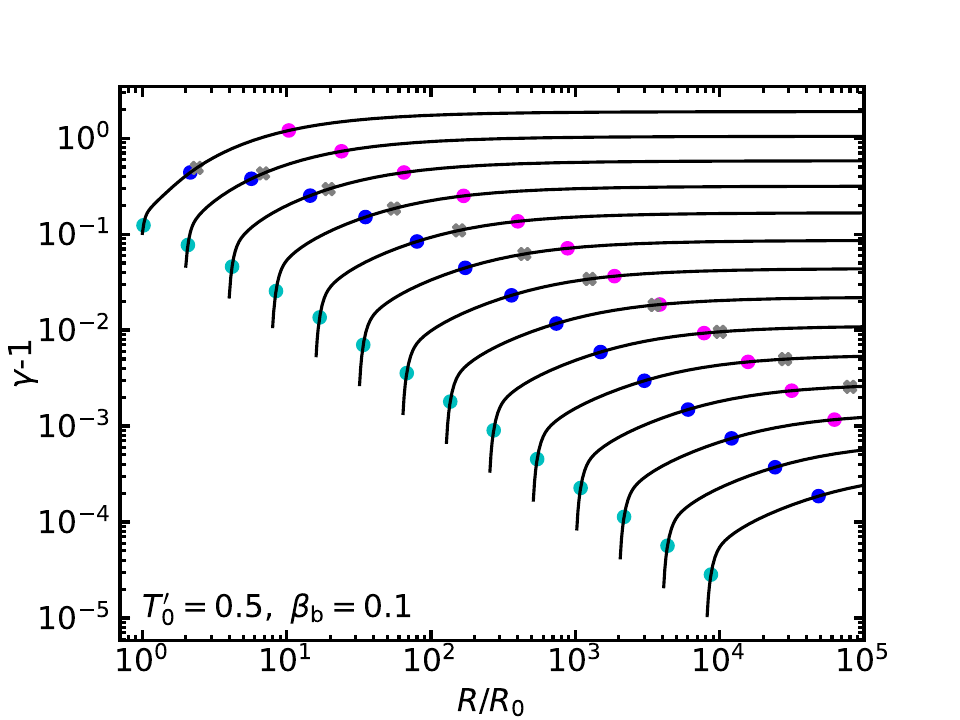}
	\includegraphics[width=0.5\textwidth]{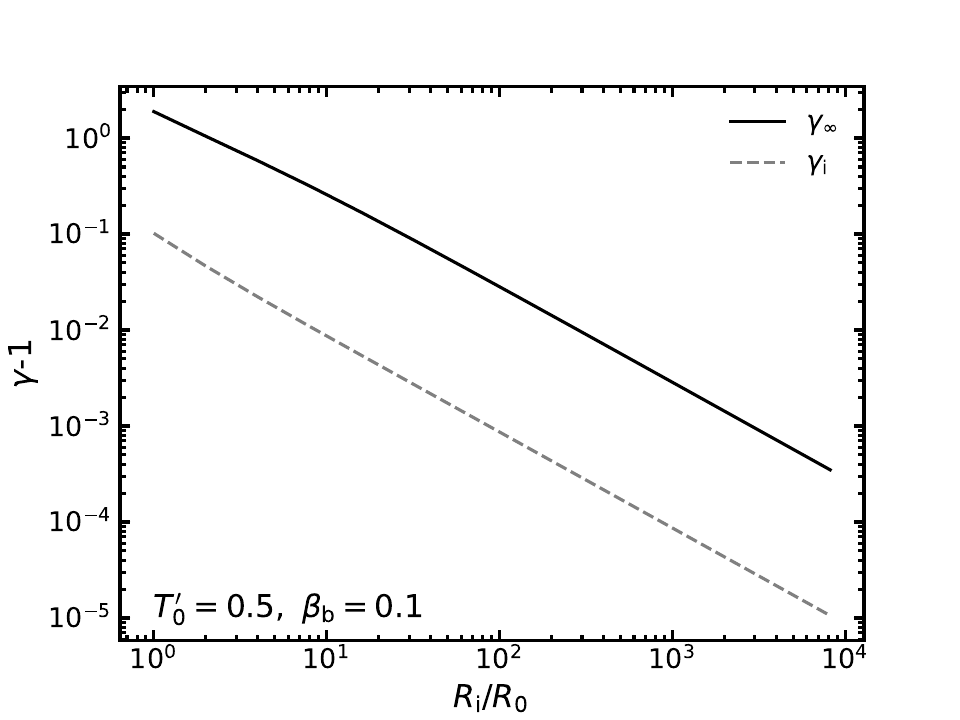}
	\caption{Same as Figure \ref{fig:Rho_dist} but plotted for the change of Lorentz factor. The right two panels show the initial and terminal values of Lorentz factor for each field line. Here, $\gamma - 1$ is used for better clarity.
		\label{fig:Gamma_dist}}
\end{figure*}

\begin{figure*}[h!t!]
	\includegraphics[width=0.5\textwidth]{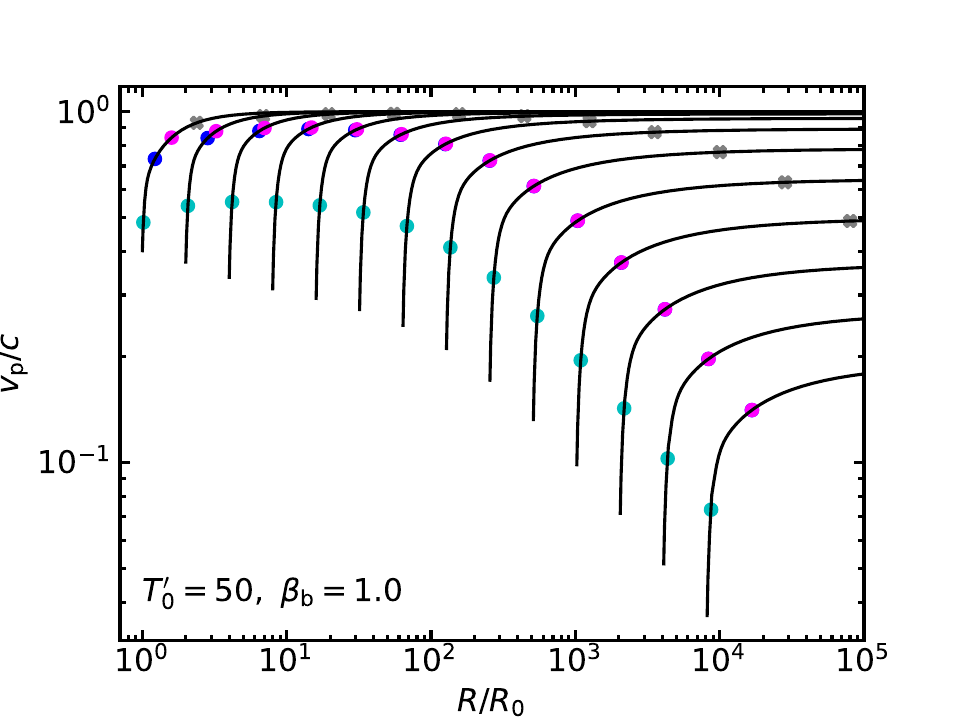}
    \includegraphics[width=0.5\textwidth]{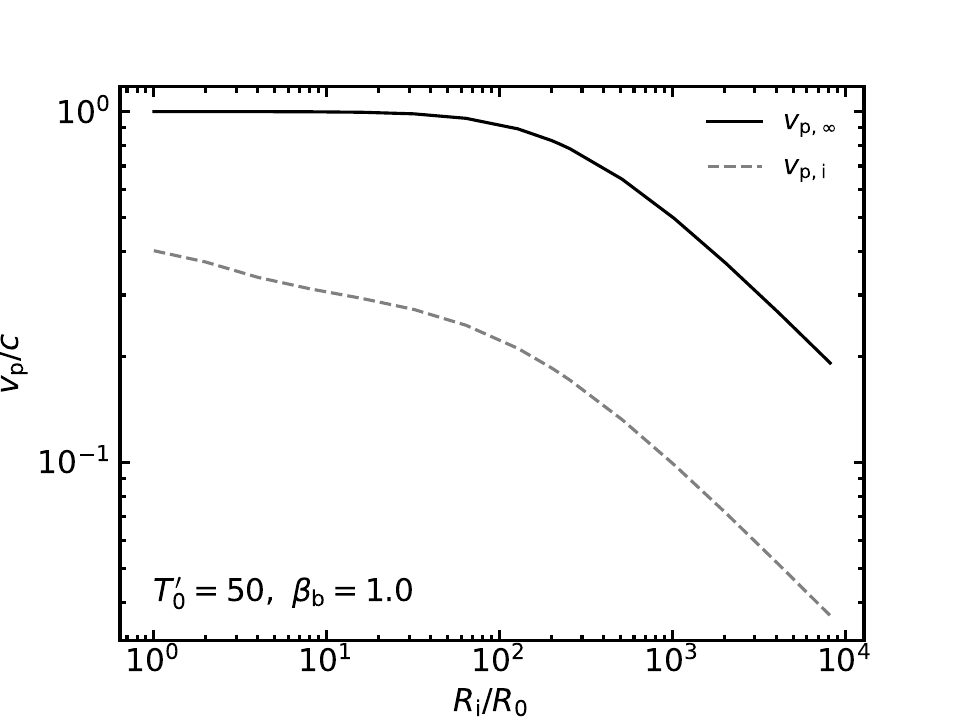} \\
	\includegraphics[width=0.5\textwidth]{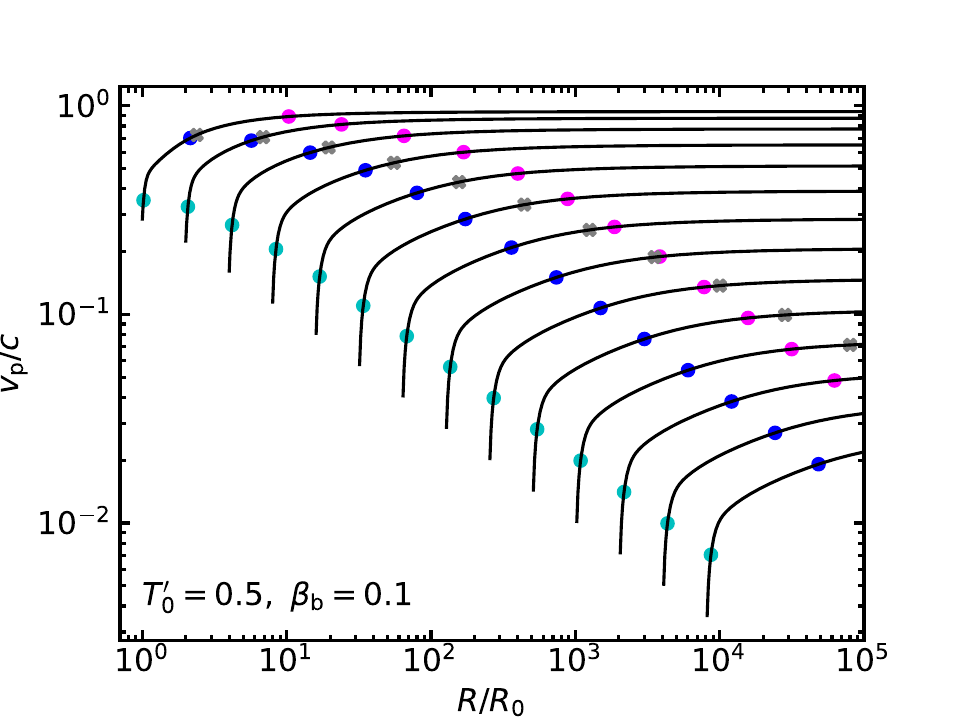}
    \includegraphics[width=0.5\textwidth]{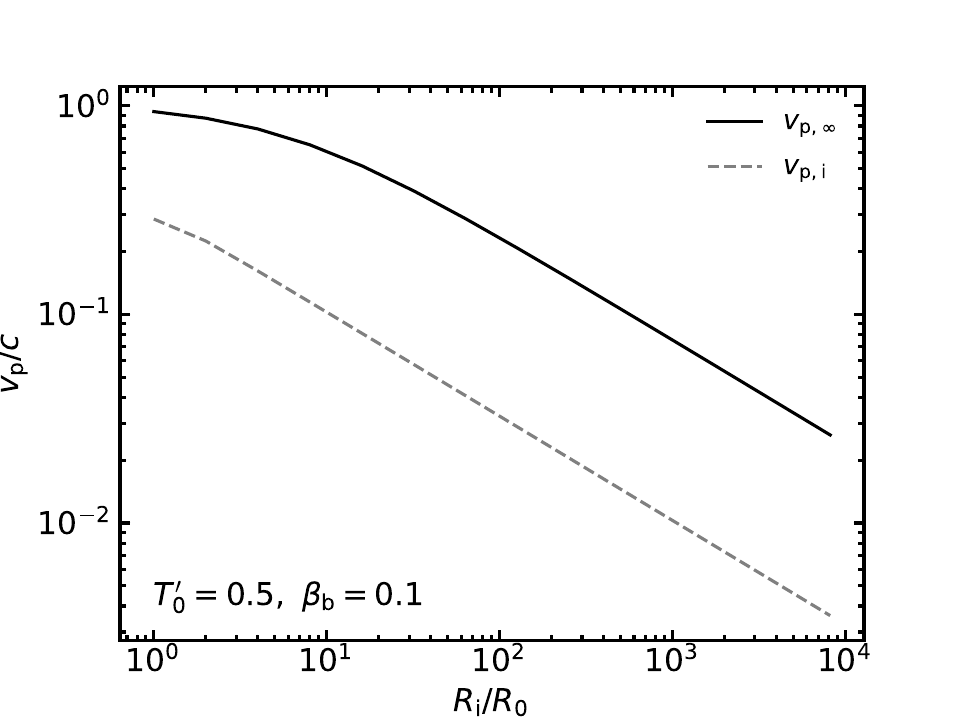}
	\caption{Same as Figure \ref{fig:Rho_dist} but plotted for the change of poloidal velocity. The right two panels show the initial and terminal values for each field line.
		\label{fig:Vp_dist}}
\end{figure*}

Figure \ref{fig:blines} shows the distribution of the critical points. For the case $T_{\mathrm i, 0}^\prime = 50$ and $\beta_\mathrm{b}=1.0$ (uppler panel), since its much higher temperature, the distances among the critical points are much closer. On the contrary, the Aflv\'en points and the F points are very far away form the S points in the case $T_{\mathrm i, 0}^\prime = 0.5$ and $\beta_\mathrm{b}=0.1$, due to its relatively lower temperature. 

\begin{figure*}[h!t!]
	\includegraphics[width=0.5\textwidth]{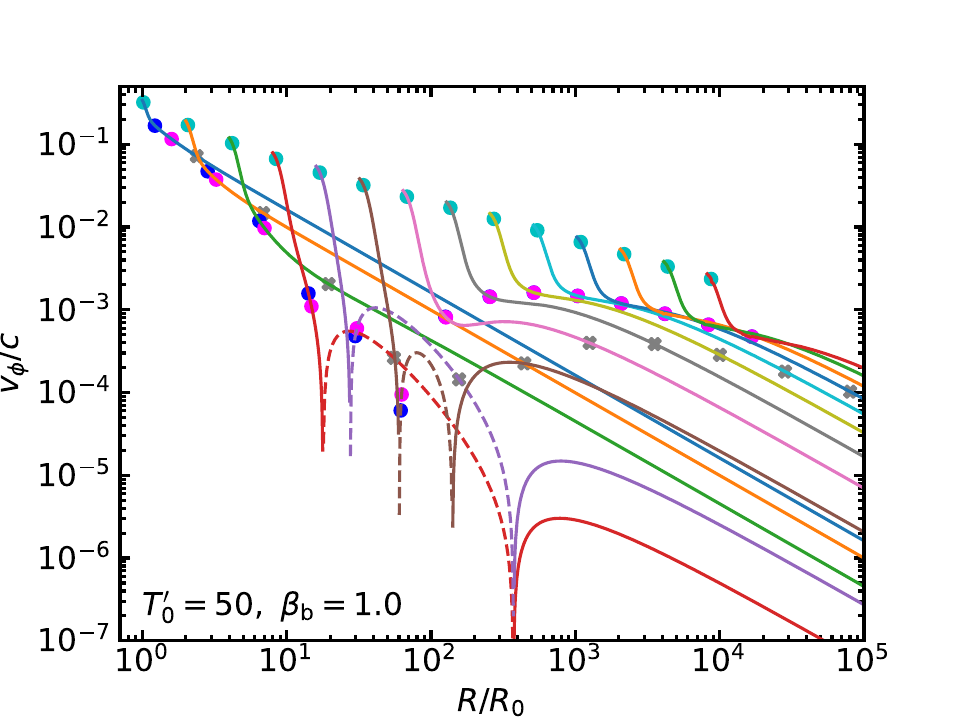}
    \includegraphics[width=0.5\textwidth]{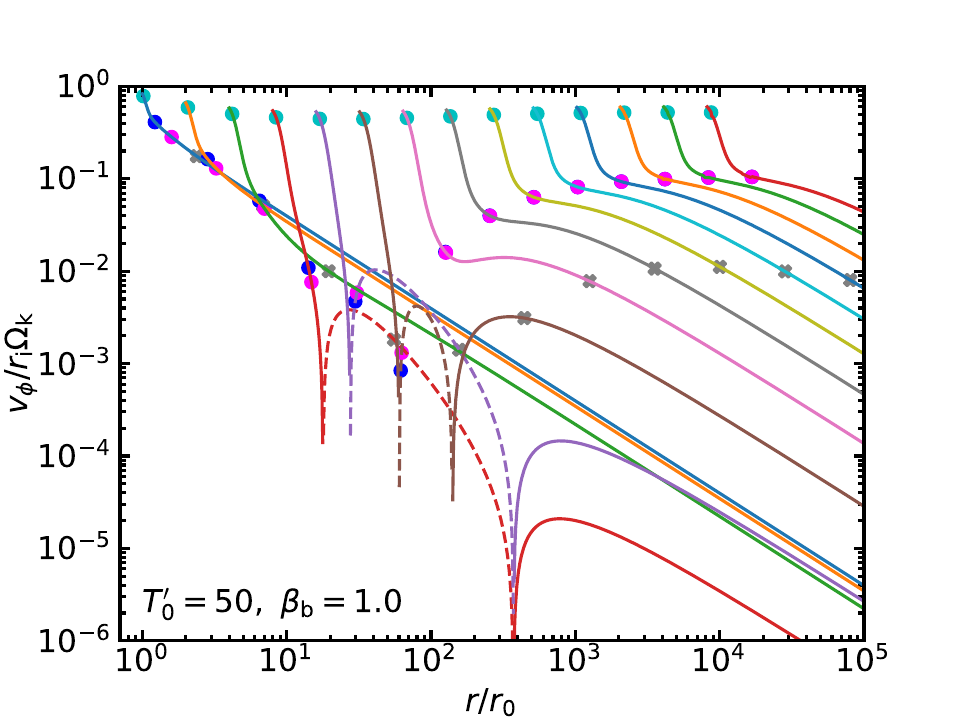}
	\includegraphics[width=0.5\textwidth]{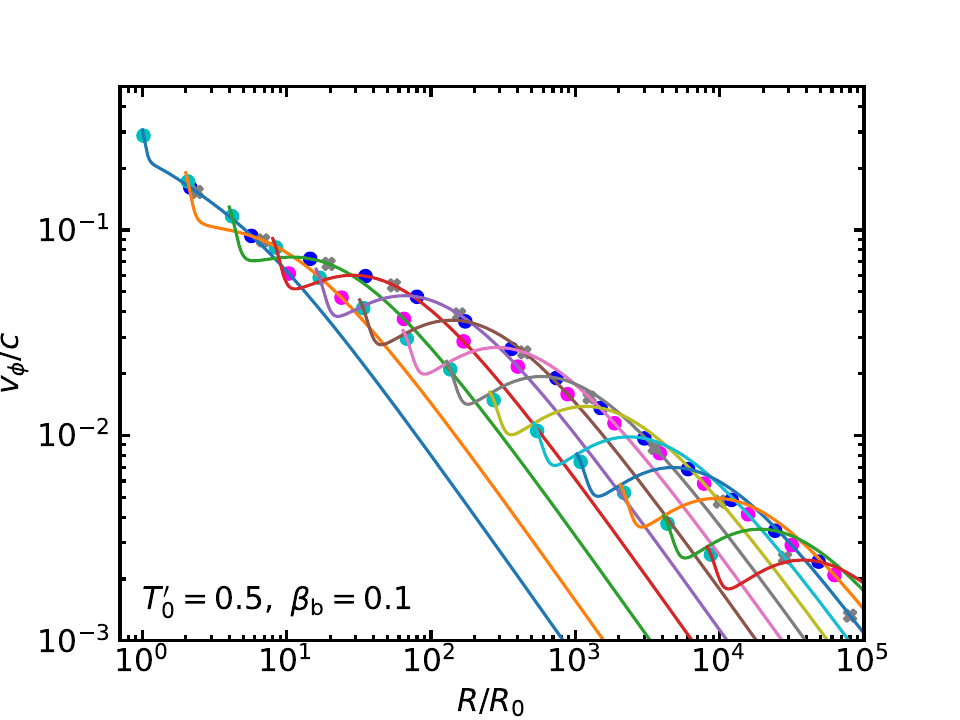}
    \includegraphics[width=0.5\textwidth]{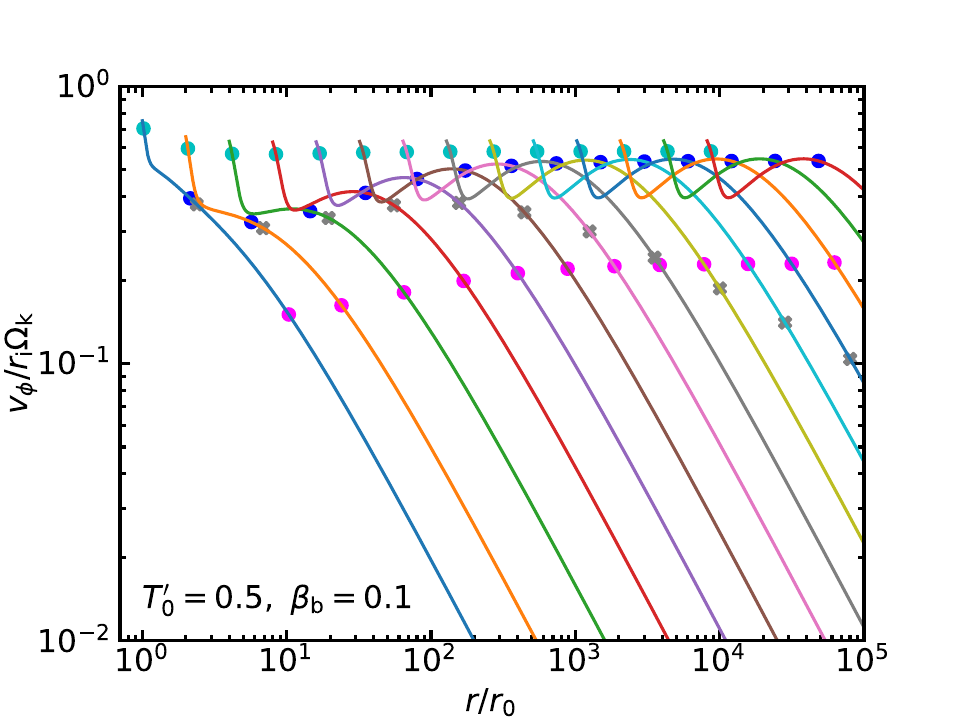}
	\caption{Same as Figure \ref{fig:Rho_dist} but plotted for the change of toroidal velocity. The left and right panels are separately plotted in units of the speed of light and Keplerian velocity. For transparency, the curves are plotted in different colors. Since the y-axis uses a log scale, the negative values of $v_\phi$ (if present) are displayed as their absolute values and distinguished by dashed lines.
	\label{fig:Vphi_dist}}
\end{figure*}

\begin{figure}[h!t!]
	\includegraphics[width=0.5\textwidth]{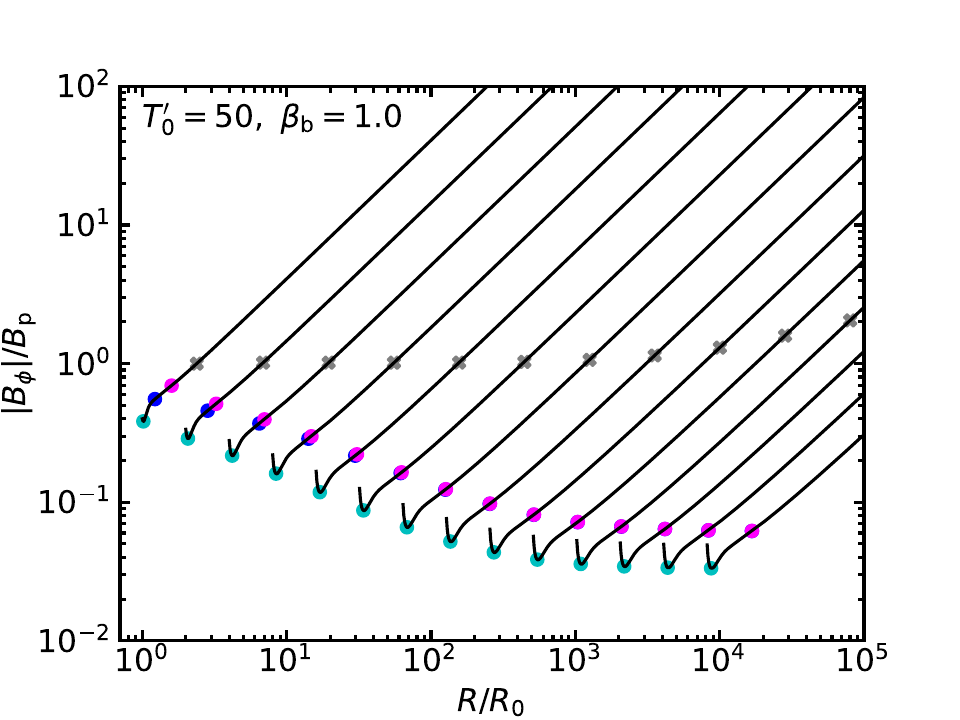}\\
	\includegraphics[width=0.5\textwidth]{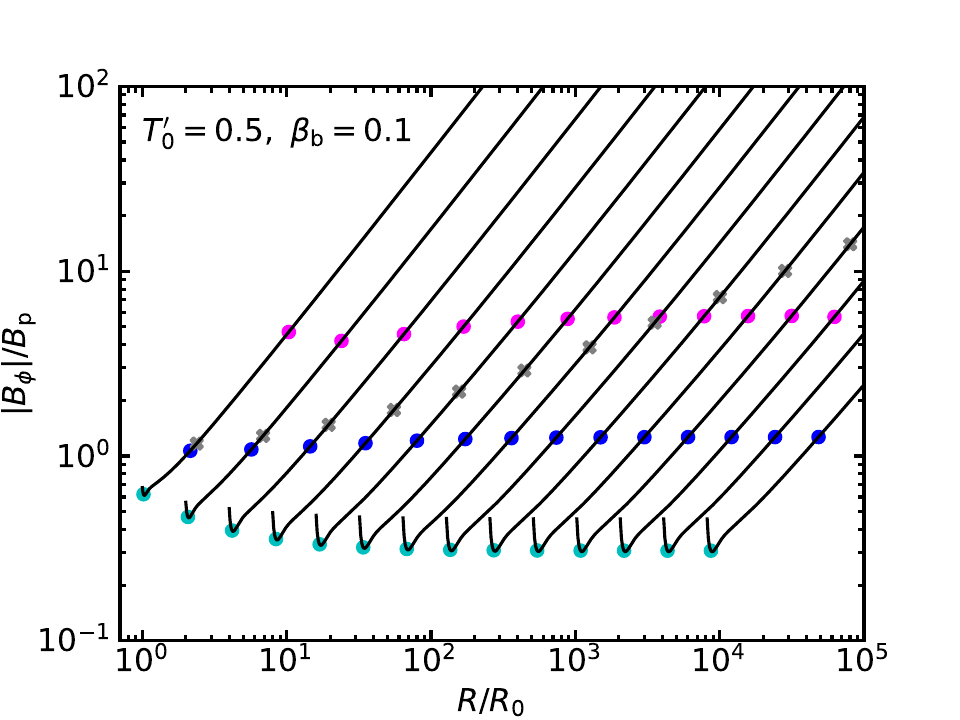}
	\caption{Same as Figure \ref{fig:Rho_dist} but plotted for the change of the ratio of toroidal to poloidal magnetic field.
		\label{fig:Bphi2Bp_dist}}
\end{figure}

Figure \ref{fig:Rho_dist}-\ref{fig:Bphi2Bp_dist} shows the distribution of different variables such as density, Lorentz factor, polodal velocity, toroidal velocity, and the ratio of toroidal to poloidal component of magnetic field. The upper(lower) panels are the solutions under the parameter $T_{\mathrm i,0}^\prime = 50$, $\beta_\mathrm{b}=1.0$($T_{\mathrm i,0}^\prime = 0.5$, $\beta_\mathrm{b}=0.1$). 

\begin{figure*}[h!t!]
    \centering
    \includegraphics[width=0.5\textwidth]{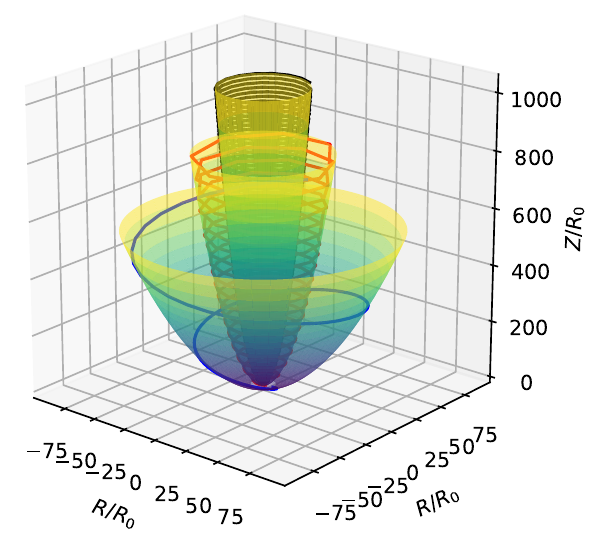}
    \includegraphics[width=0.45\textwidth]{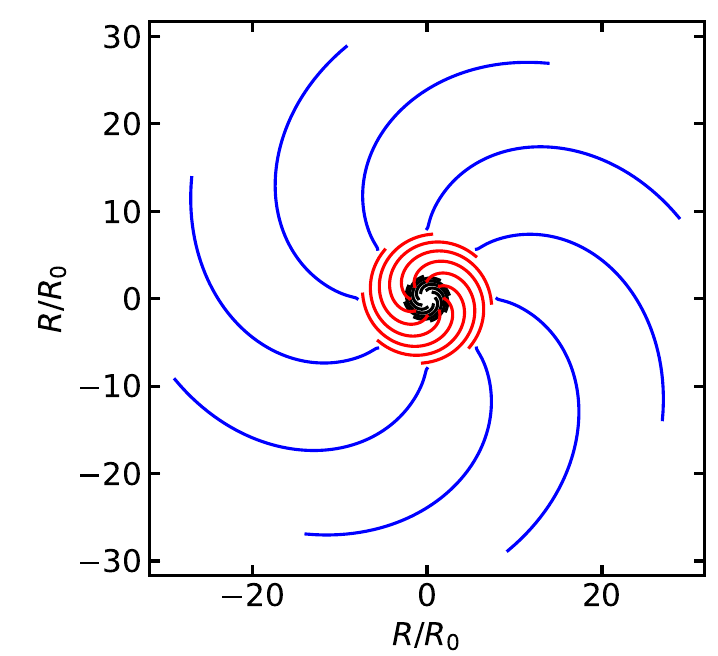}\\
    \includegraphics[width=0.5\textwidth]{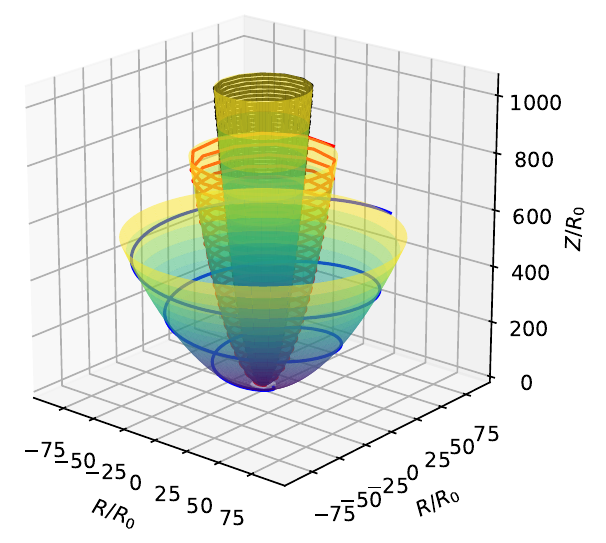}
    \includegraphics[width=0.45\textwidth]{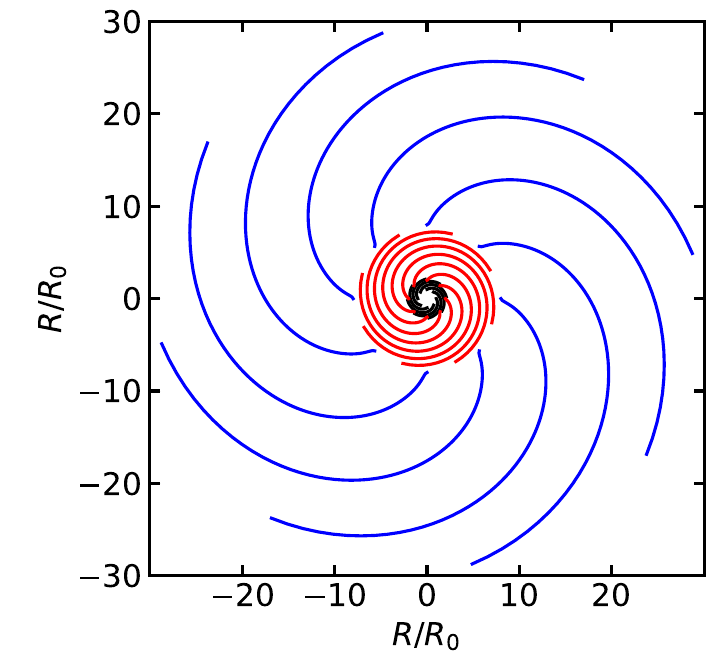}
    \caption{The 3D field lines rooted at $R_i/R_0$ = 1, 2, 8 and their projections in the disk mid-plane. The upper and lower panels separately corresponds to \{$T_0^\prime = 50$, $\beta_\mathrm{b}=1.0$\} and \{$T_0^\prime = 0.5$, $\beta_\mathrm{b}=0.1$\}
    \label{fig:blines_surf}}
\end{figure*}

It is worthwhile to note that the acceleration process are significantly different with each other for the two groups of parameters. For the case of lower temperature and lower plasma beta ($T_{\mathrm i,0}^\prime = 0.5$, $\beta_\mathrm{b}^\prime = 0.1$), the outflow is accelerated thermally up to slow magnetosonic point characterized by the density dramatically decreasing along with each field line (Figure \ref{fig:Rho_dist}), and then centrifugally driven untill Alfv\'en point, featured by the rising up of $v_\phi\propto R\Omega$ between S and A point (see the lower panel of Figure \ref{fig:Vphi_dist}). After the Alfv\'en point the inertia of matter does not allow corotation of the magnetic field. As a result, the magnetic field lines bend, developing a strong toroidal component, and the magnetic pressure gradient gradually becomes the dominant force to accelerate the flow (see the lower panel of Figure \ref{fig:Bphi2Bp_dist}). The flow is almost not accelerated after the fast point, i.e., $v_\mathrm{p} \sim \mathrm{const}$ (see the lower panel of Figure \ref{fig:Vp_dist}), while the rest physical quantities present a power-law dependences with the cylindrical radius $R$, i.e., $\rho \propto R^{-2}$ (Figure \ref{fig:Rho_dist}), $v_\phi\propto R^{-1}$ (Figure \ref{fig:Vphi_dist}), and $B_\phi/B_\mathrm{p}\propto R$ (Figure \ref{fig:Bphi2Bp_dist}). Whereas for the higher temperature and higher plasma beta ($T_{\mathrm i,0}^\prime = 50$, $\beta_\mathrm{b}^\prime = 1.0$), the outflow is mainly accelerated by thermal pressure up to fast point. The contribution of the magnetically centrifugal force is not significant, so the toroidal velocity near the Alfv\'en point is smaller than that of the lower temperature case. As a result, the toroidal magnetic field is not apparent until the light cylinder. The discrepancy on the production of toroidal component of the magnetic field between the lower temperature and higher temperature case is also illuminated in a more intuitive way in Figure \ref{fig:blines_surf}.

\begin{figure*}[h!t!]
	\includegraphics[width=0.33\textwidth]{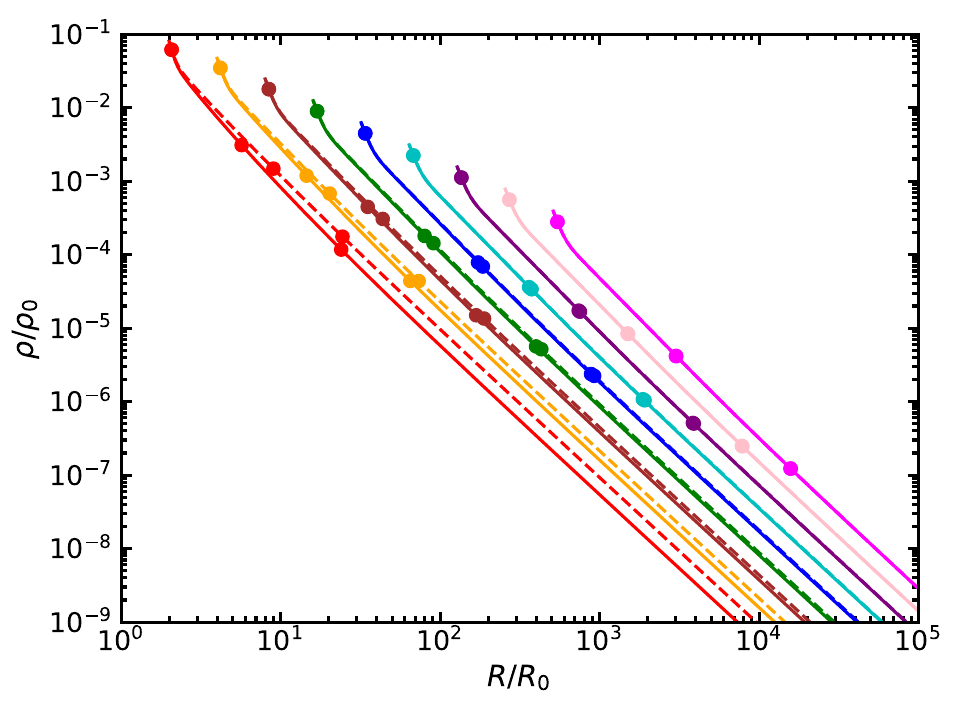}
	\includegraphics[width=0.33\textwidth]{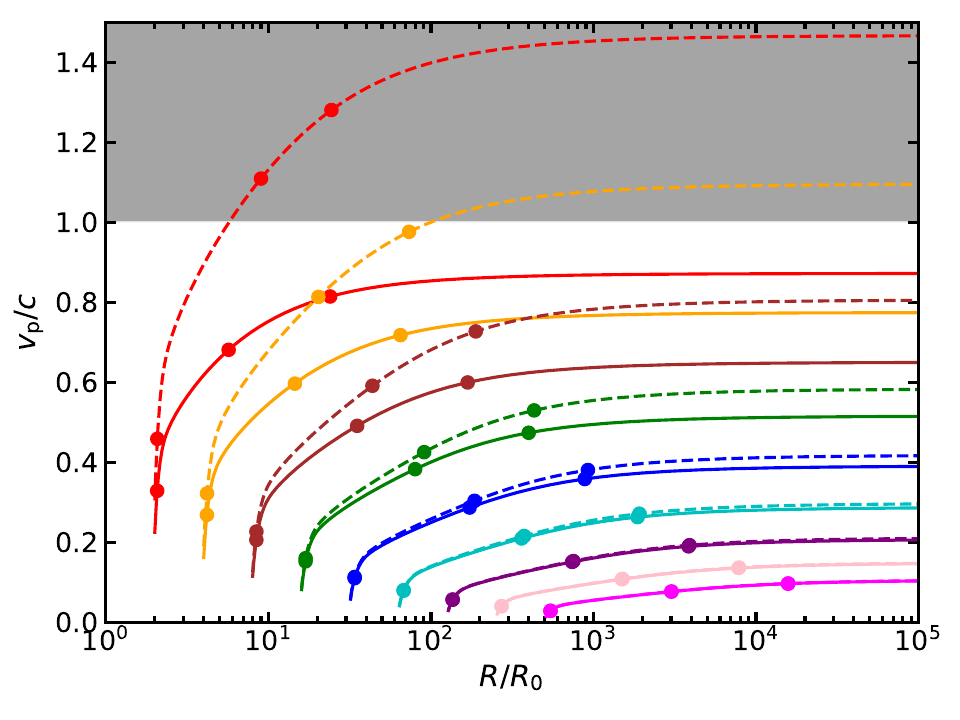}
	\includegraphics[width=0.33\textwidth]{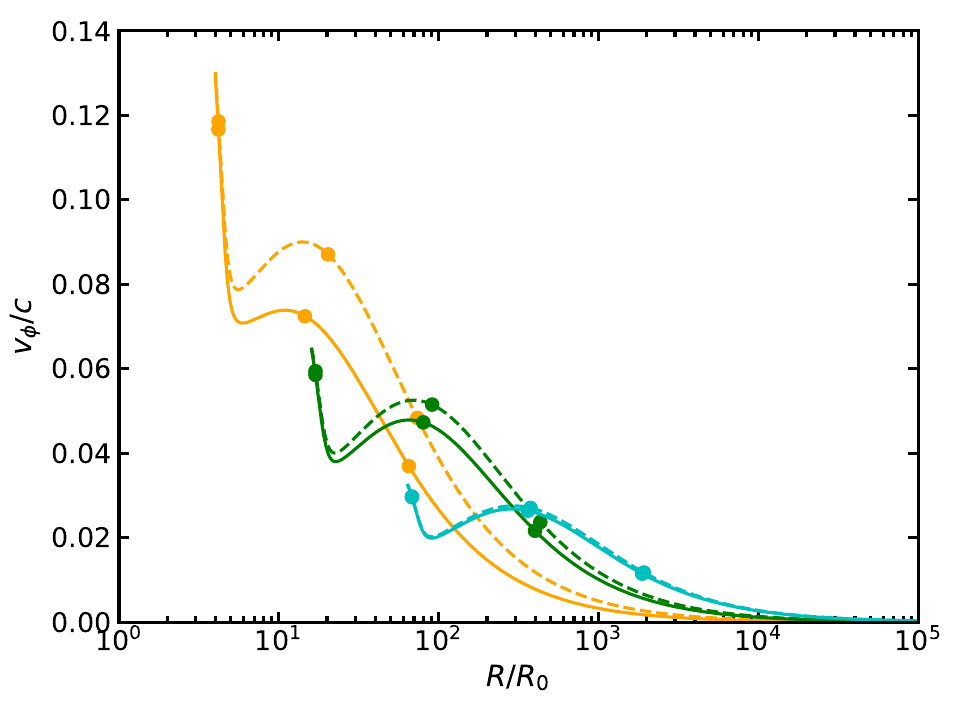}
	\caption{The change of the matter density (left panel), poloidal velocity (middle panel) and toroidal velocity (right panel) along a series of magnetic lines rooted at different disk radii, with the parameters setted as \{$T_\mathrm{0}^\prime = 0.5$, $\beta_\mathrm{b}=0.1$\}. The GR results are plotted with solid lines, while the Newtonian results dashed lines. The colored circles on each line denote the slow, Alfv\'en and fast points.
		\label{fig:GR2NT}}
\end{figure*}

\subsection{Ultra fast outflows in AGN/XRBs}

Recent surveys with XMM-Newton and Suzaku have shown that ultra fast outflows (UFOs) found through blueshifted Fe xxv and Fe xxvi  K-shell absorption lines in hard X-rays band are observed in a significant fraction ($\sim$ 40-50\%) of AGN. The outflowing velocity of the absorbers is in the range $v_\mathrm{w}\sim 0.03-0.3 c$ (e.g., \citealp{Reeves2009, Tombesi2010a, Tombesi2010b, Tombesi2011, Tombesi2013, Tombesi2015, Gofford2013, Gofford2015, Parker2017, Reeves2020, Matzeu2023, Lanzuisi2024}). The ionization parameter of UFOs spans $3 \lesssim \log \xi \lesssim 6$. The highly ionized UFOs are also observed in some X-ray binaries (XRBs; e.g., \citealp{Miller2006, Miller2008, Miller2012, King2014, Kosec2018, Wang2021}). Given the high velocity, it is believed that the UFOs originate from the accretion disk. The observed power-law hard X-ray spectra of AGNs/XRBs are most likely due to the inverse Compton scattering of soft photons by a population of hot electrons in the coronae located either above the disk or in the truncated radius of the disk. Hard X-ray observations indicate that AGN coronae have a fairly limited range of electron (lepton) temperatures imposed by efficient Compton cooling ( $T_\mathrm{e}^\prime \approx 0.03-1.36$, \citealp{Fabian2015, Fabian2017, Kamraj2022, Bambic2024}). If we take 0.5 as a fiducial value for $T_\mathrm{i}^\prime$, then comparing the results of the terminal velocity of the outflow with the observations is meaningful to evaluate the feasibility of the magneitic model. From the bottom panels of Figure \ref{fig:Vp_dist} one can see that the magnetically driven outflow under the parameters $T_\mathrm{i, 0}^\prime=0.5$ and $\beta_b=0.1$ could indeed have a typical velocity range in $\sim 0.03-0.9 c$, which is compatible with the observations. In addition, we find that the outflow velocity in the inner region of the accretion flow could be very high nearly as light-speed. Considering the collimation effect of the toroidal magnetic fields in the inner region, such inner outflow might also account for the ultra-relativistic jet of AGN/XRBs.

\subsection{Jets/Outflows of GRBs}
As shown in the top two panels of Figure \ref{fig:Gamma_dist} and \ref{fig:Vp_dist}, the terminal velocity of the magnetic-driven outflow can be extremely relativistic, the Lorentz factor could even be as high as $\sim100$, as long as $T_{\mathrm i}^\prime$ is large (e.g., $T_{\mathrm i}^\prime=50$) and $\rho_{\mathrm i}^\prime$ is small (e.g., $\rho_{\mathrm i}^\prime=0.01$). This result naturally bring us a curiosity that whether it is possible to drive the GRBs jets in this magnetic scenario? Since either the prompt or the afterglow emission are originated from the jet itself or its interaction with the circum medium, it is hardly to directly obtain the information of the central disk about the its temperature, the density and the magnetic strength by means of the observations. Nontheless, a hyper-accretion disk is believed to be the central engine of GRBs, the typical temperature and density of such a disk is $T\sim10^{11} K$ and $\rho \sim 10^{10} \mathrm{g\ cm^{-3}}$, and the corresponding dimensionless temperature is about $T_\mathrm{i}^\prime =k_\mathrm{B} T/m_\mathrm{e}c^2 \sim 20$, this value is at the same order of magnitude which is used in our calculation. In addition, there might be a hot corona above the hyperaccreting disk \citep{RS05}, and the termperature of the corona can be significantly larger than that of the disk \citep{Kawabata2008}. As a result, it seems possible that the magnetic-driven outflow from a hyperaccreting disk account for the GRB jets/outflows.

We have clarified that relativistic outflows could originate from current magnetic acceleration models, albeit with parameters (especially temperature) that may sometimes be rather extreme. However, the actual scenario could be more intricate. As mentioned previously, once the outflow reaches the Alfv\'en point (or the light cylinder), the toroidal component of the magnetic field becomes dominant, and acceleration via magnetic pressure gradients becomes limited. In fact, at this juncture, the outflow is susceptible to kink instability (e.g., \citealt{GS2006}) or magnetic reconnection (if the magnetic field is not axisymmetric, e.g., \citealt{Drenkhahn2002}), leading to the dissipation of Poynting energy. These kinds of dissipation not only generate electromagnetic radiation but also further augment the magnetic pressure gradient, allowing the outflow to continue accelerating beyond the Alfv\'en point (or the light cylinder). If this is indeed the case, one might expect that, under conditions allowing outflow, the temperatures and magnetization would be more moderate than their current values..

\section{Summary}\label{sec:summary}
Based on general relativistic MHD equations, we establish a formulation to describe the outflows driven by large-scale magnetic fields from disks in Schwarzschild spacetime. The outflow solution manifests as a contour level of a ``Bernoulli" function, which is determined by ensuring that it passes through both the slow and fast magnetosonic points. This approach is a general relativistic extension to the classical treatment of Cao and Spruit (1994). The existence of magnetically driven outflow solutions necessitates a constrained initial plasma beta within the range of approximately $0.02 \lesssim \beta_\mathrm{b} \lesssim 2$; outside this range, outflows may be generated by different physical processes. The terminal velocity of the outflows is directly proportional to the initial temperature and inversely proportional to the initial relative density (or directly proportional to the magnetization). Under reasonable parameters, relativistic outflows and jets from AGNs, XRBs, and GRBs could be driven by large-scale magnetic fields, provided they possess high initial magnetization and initial temperature. Considering the effects of black hole spin and various thermodynamic processes will be important in the future for deepening our understanding of accretion processes and jet formation mechanisms in various astrophysical objects.

\begin{acknowledgments}

We thank the anonymous referee for constructive comments and suggestions. This work was performed under the auspices of the Science and Technology Foundation of Guizhou Province (grant No. QianKeHeJiChu ZK[2021]027), the Special Foundation for Theoretical Physics Research Program of National Natural Science Foundation of China (grant No. 11847102), the National Key R\&D Program of China (grant No. 2020YFC2201400), and the National Natural Science Foundation of China (grant No. 12473012).
\end{acknowledgments}

%

\vspace{5mm}





\appendix

\section{Domain of definition of the Bernoulli function}\label{sec:domainXY}
The Bernoulli function $\tilde{H}(x,y)$ can be well defined only as $\mathcal{D}(x,y)>0$ which gives the following condition:

\begin{equation}\label{eq:eqY}
  A(x) Y^2 - 2Y +C(x)>0,
\end{equation}

\noindent where we have introduced two auxiliary functions:

\begin{equation}
	Y(y; y_\mathrm{i}, T_\mathrm{i}^\prime, \gamma) = \frac{\tilde{h}_\mathrm{A} y }{\tilde{h}(y)},
\end{equation}

\begin{equation}
	A(x; x0, m, \mathcal{R}_\mathrm{i}, g) = \frac{\alpha^2-\omega x^2}{\alpha_\mathrm{A}^2-\omega},
\end{equation}

\begin{equation}
	C(x; x0, m, \mathcal{R}_\mathrm{i}, g) = \frac{-\alpha^2\omega + \alpha_\mathrm{A}^4x^2}{(\alpha_\mathrm{A}^2-\omega)\alpha^2 x^2}.
\end{equation}

The two parameters $m$ and $\gamma$ are set to fixed values for the whole system. For a certain field line originating at $\mathcal{R}_\mathrm{i}$ with its geometry marked by $g$, given a prescribed value of $T_\mathrm{i}^\prime$, the function $Y(y)$ depends on $y_\mathrm{i}$, whereas the function $A(x)$ and $C(x)$ depend on $x_0$. The function $Y(y)$ is strictly increasing from $Y(0)=0$ to $Y(+\infty)=+\infty$ with $Y(1)=1$. For each $x$, \eqref{eq:eqY} has 1 or 2 positive roots of $Y(y)$, delimiting the permitted domain of $y$ so that $\tilde{H}(x,y)$ could be well defined. An interesting aspect to note here is that the complete domain on the $x$-$y$ plane could not be definitively determined until a solution encompassing $x_0$ and $y_i$ is obtained. This contrasts with the approach in \cite{DaigneDrenkhahn2002}, therein the outflow from the surface of a rotating neutron star was analyzed exclusively in the equatorial plane. This simpler framework allowed them to define the entire $x$ and $y$ domain at the outset of their solution. This discrepancy arises from the fact that the motion in the $R$-$Z$ plane involves an arbitrary configuration of the poloidal magnetic field, rather than an simple one which has a constant opening angle of the flux tube.

\section{Comparision between the general relativistic and Newtonian results}

In this section, we compare the classical and GR theory about the disk outflow. The classical theory is referred to \cite{CaoSpruit1994}. We keep their normalization rule where the specific energy is normalized with $GM/R_\mathrm{A}$, and the dimensionless Bernoulli function is written as 

\begin{equation}
\begin{aligned}
	\tilde{H}(x,y) & =\frac{\beta}{2y^{2}}\left(\frac{B_\mathrm{p}}{B_{\mathrm{Ap}}}\right)^{2}+\frac{1}{2}\omega\left[\frac{x^{2}-1}{x(1-y)}\right]^{2} \\
	& + \frac{\Theta_\mathrm{i}}{\Gamma-1}\left(\frac{y}{y_\mathrm{i}}\right)^{\Gamma-1} - \frac{1}{\sqrt{x^2+\tilde{z}^2 x_0^2}} -\frac{\omega x^2}{2},
\end{aligned}
\end{equation}

%
%
%

\noindent one difference is the third term, here we adopt the adiabatic expression for the specific enthalpy. In addition, $\tilde z$ denotes the dimensionless height in unit of $R_0$, i.e., $\tilde z = Z/R_0$, and the quantities $\omega$, $\beta$ and $\Theta_{\mathrm i}$ are expressed as 
\begin{equation}
	\begin{aligned}
	&\omega=\frac{\Omega^{2}R_{\mathrm{A}}^{3}}{G{M_*}} = \frac{1}{x_\mathrm{i}^3} = \frac{1}{(\tilde r_\mathrm{i} x_0)^3}, \\
	&\beta=\frac{B_\mathrm{Ap}^{2}R_{\mathrm{A}}}{4\pi \rho_\mathrm{A} G{M_*}} = \left(\frac{B_\mathrm{Ap}}{B_\mathrm{pi}}\right)^2\frac{y_\mathrm{i}}{x_0 \rho_\mathrm{i}^{\prime\prime}}, \\
	&\Theta_\mathrm{i} = \frac{c_\mathrm{s, i}^2 R_\mathrm{A}}{G {M_*}} = \frac{T_\mathrm{i}^{\prime\prime}}{x_0},
	\end{aligned}
\end{equation}

\noindent here the parameter $\rho_\mathrm{i}^{\prime\prime}$ and $T_\mathrm{i}^{\prime\prime}$ is defined as 

\begin{equation}
	\rho_\mathrm{i}^{\prime\prime} = \frac{4\pi \rho_\mathrm{i}}{B_\mathrm{pi}^2} \frac{G {M_*}}{R_0},
\end{equation}

\begin{equation}
	T_\mathrm{i}^{\prime\prime} = \frac{c_\mathrm{s, i}^2 R_0}{G {M_*}}.
\end{equation}

\noindent Comparing with the definitions in equation \eqref{eq:rho_ip} and \eqref{eq:T_ip}, one gets the relation between the two normalization rules about the initial outflow temperature and density, i.e., 

\begin{equation}
	\frac{\rho_\mathrm{i}^{\prime\prime}}{\rho_\mathrm{i}^\prime} = \frac{G{M_*}}{R_0 c^2} = \frac{R_\mathrm{g}}{R_0} = m,
\end{equation}

\begin{equation}
	\frac{T_\mathrm{i}^{\prime\prime}}{T_\mathrm{i}^\prime} = \frac {R_0 c^2}{G{M_*}} = \frac{R_0}{R_\mathrm{g}}  = \frac{1}{m}.
\end{equation}

It is easy to get the expressions of $v_\mathrm{p}$ and $v_\phi$ as follow, 

\begin{equation}
	v_\mathrm{p}/c = \sqrt{\frac{m y_\mathrm{i}}{\rho_{\mathrm i}^{\prime\prime}}} \frac{1}{y} \frac{B_\mathrm{p}}{B_\mathrm{pi}},
\end{equation}

\begin{equation}
	v_\phi/c = \sqrt{\frac{m}{\tilde r_\mathrm{i}^3}} \frac{xy-1/x}{x_0(y-1)}.
\end{equation}

As an example for comparision, we calculate the distribution of the outflow separately in Newtonian and GR theory. The parameters are setted as the same with that in section \ref{subsec: distribution}, i.e., $T_\mathrm{i}^\prime = T_\mathrm{0}^\prime/\tilde{r}_i$ and $\rho_\mathrm{i}^\prime = \beta_\mathrm{b}/2T_\mathrm{i}^\prime$, with $T_\mathrm{0}^\prime = 0.5$ and $\beta_\mathrm{b}=0.1$. The results are shown in Figure \ref{fig:GR2NT}, and we can find that the results of Newtonian and GR are almost the same in the outer side of the disk, since the gravity is weaker and the velocity is smaller and consequently the relativistic effects is not important. However, the two results  gradually deviate with each other when the foot point moves toward the inner side of the disk, since the gravity is stronger and the velocity is larger. Especially, when the foot point is enough close to the black hole and the initial temperature is enough high, the ultimate poloidal velocity in Newtonian result can even be larger than light speed, it is obviously unrealistic and the relativistic effect must be considered in this case.  


\bibliography{MagneticOutflow}{}
\bibliographystyle{aasjournal}


\end{document}